\documentclass[fleqn,usenatbib]{mnras}

\usepackage[T1]{fontenc}
\usepackage{ae,aecompl}
\usepackage{caption}
\usepackage{subcaption}
\usepackage{xcolor}
\usepackage{tablefootnote}

\usepackage{graphicx}	
\usepackage{amsmath}	
\usepackage{amssymb}	

\newcommand{\vsini}{$v$\,sin\,$i$}
\newcommand{\dmppF}{\hbox{DMPP-4}}

\title[DMPP-4]{DMPP-4: Candidate sub-Neptune mass planets orbiting a naked-eye star}

\author[J.R. Barnes et al.]
{J.R.~Barnes$^{1}$,
{ M.R.~Standing$^{1}$},
C.A.~Haswell$^{1}$,
D.~Staab$^{1,2}$,
J.P.J.~Doherty$^{1}$,\newauthor
M. Waller-Bridge$^{1}$,
L.~Fossati$^{3}$,
M.~Soto$^{4}$,
G.Anglada-Escud\'{e}$^{5,6}$,
J.Llama$^{7}$, \newauthor
C.~McCune$^{1}$,
{ F.W.~Lewis$^{8}$}
\\
$^{1}$ School of Physical Sciences, The Open University, Walton Hall, Milton Keynes. MK7 6AA. UK \\
$^{2}$ AVS, Rutherford Appleton Laboratory, Harwell, Oxford, OX11 0QX United Kingdom \\
$^{3}$ Space Research Institute, Austrian Academy of Sciences, Schmiedlstrasse 6, 8042 Graz, Austria \\
$^{4}$ School of Physics and Astronomy, Queen Mary, University of London, 327 Mile End Rd, London. UK \\
$^{5}$ Institut de Ciencies de l’Espai (ICE, CSIC), Campus UAB, Can Magrans s/n, 08193 Bellaterra, Spain \\
$^{6}$ Institut d’Estudis Espacials de Catalunya (IEEC), 08034 Barcelona, Spain \\
$^{7}$ Lowell Observatory, 1400 West Mars Hill Rd
Flagstaff AZ, 86001, USA \\
$^{8}$ Faulkes Telescope Project, School of Physics and Astronomy, Queen's Building, The Parade, Cardiff. CF24 3AA. Wales. UK
}
\date{Accepted for publication in MNRAS, July 2023.}

\pubyear{2022}

\begin{document}
\label{firstpage}
\maketitle

\begin{abstract}
We present radial velocity measurements of the very bright ($V\sim5.7$) nearby { F} star, DMPP-4 (HD~184960). { The anomalously low Ca~{\sc ii}~H\&K emission suggests mass loss from planets orbiting a low activity host star. Periodic radial velocity variability with $\sim 10$ m\,s$^{-1}$ amplitude is found to persist over a $>4$ year timescale. Although the non-simultaneous photometric variability in four {\sc tess} sectors supports the view of an inactive star, we identify periodic photometric signals and also find spectroscopic evidence for stellar activity. We used a posterior sampling algorithm that includes the number of Keplerian signals, $N_\textrm{p}$, as a free parameter to test and compare (1) purely Keplerian models (2) a Keplerian model with linear activity correlation and (3) Keplerian models with Gaussian processes. A preferred model, with one Keplerian and quasi-periodic Gaussian process indicates a planet with a period of $P_\textrm{b} = 3.4982^{+0.0015}_{-0.0027}$~d and corresponding minimum mass of $m_\textrm{b}\,\textrm{sin}\,i = 12.2^{+1.8}_{-1.9}$~M$_\oplus$. Without further high time resolution observations over a longer timescale, we cannot definitively rule out the purely Keplerian model with 2 candidates planets with $P_\textrm{b} = 2.4570^{+0.0026}_{-0.0462}$~d, minimum mass $m_\textrm{b}\,\textrm{sin}\,i = 8.0^{+1.1}_{-1.5}$~M$_\oplus$ and $P_\textrm{c} = 5.4196^{+0.6766}_{-0.0030}$~d and corresponding minimum mass of $m_\textrm{b}\,\textrm{sin}\,i = 12.2^{+1.4}_{-1.6}$~M$_\oplus$.  The candidate planets lie in the region below the lower-envelope of the Neptune Desert. Continued mass loss may originate from the highly irradiated planets or from an as yet undetected body in the system.}
\end{abstract}




\section{Introduction}


The dearth of planets, now commonly dubbed the ``Neptune Desert'', is a very clear region { in the planetary mass vs orbital period plane \citep{mazeh16}}. It is delineated by a sharp upper edge with a planet mass inversely proportional to the orbital period, the trend first noted by \cite{mazeh05}, and a lower boundary with masses that are roughly linearly proportional to the period. {Several} mechanisms to explain both boundaries were discussed by \cite{mazeh16} and more recently by \citet{vissapragada22}. The upper boundary can be explained as a death line where inward migrating planets lose much of their mass due to insolation from the host star. They consequently move down below the lower boundary of the desert where a large number of transiting Kepler planets were found in a ridge in the planet radius vs orbital period diagram \citep{mazeh16}. The lack of intermediate mass planets at short periods exists because the timescale for this mass-loss is thought to be short. A number of planets have nevertheless been found in the Neptune Desert \citep{west19,jenkins20,diaz20,armstrong20,burt20,jordan20,dreizler20,smith21,murgas21,kanodia21,mori22}{. These} planets orbit stars with $11.1 < V < 17.0$, except for a single example with $V = 9.8$ \citep{jenkins20}. {Identification of planetary systems with brighter stellar hosts that are either in, or have potentially transitioned the Neptune Desert, would be advantageous for follow-up characterisation.}

{Chromospheric emission has been found to be depressed in stars harbouring mass-losing planets \citep{vidalmadjar03,haswell12wasp12}. The Ca {\sc ii} H\&K emission in these stars is below the basal level for inactive stars owing to absorbing gas that is lost from close-orbiting planets \citep{haswell20dmpp}. Hence, {\em a priori} identification of stars that harbour hot ablating planets is possible.} The Dispersed Matter Planet Project (DMPP) targets bright stars within 100\,pc, with typical {apparent magnitudes of $m_v<10$}, making follow-up characterisation easier \citep{haswell20dmpp}. {A total of 39 host stars} with anomalously low Ca II H\&K emission have been carefully selected by DMPP.  Because this gas is concentrated in the orbital plane of a planet \citep{debrecht18}, the method preferentially picks out edge-on systems. Consequently, a significantly higher than average fraction of these planets are expected to transit. Most of the DMPP planets reported to date lie just below the lower Neptune desert boundary \citep{haswell20dmpp,barnes20dmpp3,staab20dmpp1}. One explanation is that they are the stripped cores of more massive planets that have crossed the Neptune Desert. It is possible that the DMPP planets identified so far are part of the same population as the dense concentration or ``ridge'' of Kepler planets in the period-radius diagram \citep{mazeh16}. The DMPP planets thus offer valuable opportunities to understand the mechanisms sculpting exoplanet demographics, and thus the evolution of individual hot planets.

The compact multiplanet system, DMPP-1, comprises some of the most irradiated rocky planets yet known \citep{staab20dmpp1}. The DMPP planets are not typically mass-losing hot Jupiters; rather, it is likely that the quenching of Ca~{\sc ii}~H\&K emission comes from the low-mass, potentially rocky planets, or from additional planets that are below the radial velocity (RV) detection threshold. \cite{jones20dmpp1} identified a candidate transiting signature in {\sc tess} observations of DMPP-1 that would correspond to such a low-mass planet with an extended mass-losing atmosphere. This planet was not identified in the RVs reported by \cite{staab20dmpp1}. DMPP target stars could thus host analogues and precursors of catastrophically disintegrating planets \citep{rappaport12,rappaport14,sanchisojeda15k222}.

DMPP has proved very successful, with an effectively 100$\%$ success rate \citep{haswell20dmpp} on targets with sufficient observations. While most targets are in the southern hemisphere, the DMPP sample comprises a number of northern targets. Here we present SOPHIE and HARPS-N RV observations of our brightest target, the late-F star, \dmppF, which is visible to the naked eye. We present system parameters for \dmppF~in \S \ref{section:parameters}. In \S \ref{section:photometry}, we discuss ground-based and {\sc tess} photometric observations and search for evidence of the stellar rotation period. Spectroscopic activity indicators are analysed in {\S \ref{section:spectroscopic} before searching for Keplerian signals in the RVs} in \S \ref{section:RVanalysis}. A summary and further discussion {in \S \ref{section:summary} is followed by a brief conclusion in \S \ref{section:conclusion}}.

\section{System parameters of DMPP-4}
\protect\label{section:parameters}
At $V \sim 5.7$, \dmppF~is visible to the naked eye. It is a northern hemisphere star (J2000 coordinates 19:34:19.8, +51:14:11.8). Stellar observed and derived parameters are listed in Table \ref{tab:HD184960params}. The Gaia Data Release 3 parallax is $\pi = 39.323$~mas, from which a directly estimated distance ($d = 1/\pi$) of $25.43$~pc is implied\footnote{The widely reported systematics in Gaia parallaxes (e.g. see \citealt{lindegren21gaia}) would have a small effect for nearby objects.}.

A range of spectral types between F5 \citep{roman49} and F8V \citep{eggen60} have been estimated\footnote{CDS database \url{https://simbad.u-strasbg.fr}}. Table \ref{tab:HD184960params} includes estimates of $V$ and also $\textrm{B-V} = 0.4344 \pm 0.0076$ derived from photometric conversions\footnote{Gaia DR3 documentation \url{https://www.cosmos.esa.int/web/gaia-users/archive/gdr3-documentation}} documented in Gaia Data Release~3 \citep{gaia16,gaia23dr3}, which is more consistent with an earlier spectral type of F5V \citep{pecaut12,pecaut13}. Our own estimates of $T_\textrm{eff}$, $M_*$ and $T_*$ using our HARPS-N observations and {\sc species} \citep{soto18species} are also more consistent with an earlier spectral type classification of F5V. We also find an age of $2.15$~Gyr using {\sc species}. With careful consideration of macroturbulence, $v_{\rm mac}$, the equatorial rotation velocity, \vsini{} and $v_{\rm mac}$ have been obtained following the method described in \cite{murphy16}. Macroturbulence contributes significantly to the broadening, with $v_{\rm mac} = 7.0 \pm 2.0$\,kms$^{-1}$. The derived \vsini{} $=7.5^{+0.5}_{-2.0}$\,kms$^{-1}$ is consequently significantly lower than would be measured without including $v_{\rm mac}$.

\begin{table}
	\centering
  \caption[\dmppF stellar parameters]{\dmppF~stellar parameters. (1) CDS database (2) \cite{roman49} (3) \cite{eggen60} (4) Gaia Data Release 3 \citep{gaia23dr3} (5) \cite{stassun18tess} (6) \cite{pace13} (7) \cite{soto18species} (8) \cite{murphy16}.}
  \protect\label{tab:HD184960params}

  \begin{tabular}{lcc}

  \hline
   Parameter                     & value               &  Reference    \\
  \hline
   Spectral Type                 & F5 - F8V                   & 1,2,3 \\

   Mean G [mag]             & 5.5974 $\pm$ 0.0005    & 4 \\
   Mean BP [mag]            & 5.8437 $\pm$ 0.0009    & 4 \\
   Mean RP [mag]            & 5.1896 $\pm$ 0.0024    & 4 \\
   V [mag]                  & 5.7034 $\pm$ 0.0302    &  this work, from 4 \\
   B-V [mag]                & 0.4344 $\pm$ 0.0076    &  this work, from 4 \\
   {\sc tess} [mag]                    & 5.22                  & 5 \\
   $\pi$ [mas]                   & 39.323 $\pm$ 0.028    & this work, from 4 \\
   $1/\pi$ [pc]                  & 25.43 $\pm$ 0.02      & this work, from 4 \\
   log($R^{\prime}_{\rm HK}$)    & -5.24                 & 6 \\
   $T_{\rm eff}$ [K]             & 6400 $\pm$ 50         & this work, 7 \\
   Fe/H                          & -0.007 $\pm$ 0.056    & this work, 7  \\
   $\log{g}$ [cms$^{-2}$]      & 4.24 $\pm$ 0.10         & this work, 7 \\
   $v\sin{i}$ [kms$^{-1}$]     & 7.5$^{+0.5}_{-2.0}$     & this work, 8 \\
   $v_{\rm mac}$ [kms$^{-1}$]  & 7.0 $\pm$ 2.0           & this work, 8 \\
   $M_{*}$ [M$_{\sun}$]          & 1.25 $\pm$ 0.02       & this work, 7 \\
   $R_{*}$ [R$_{\sun}$]          & 1.38 $\pm$ 0.01       & this work, 7 \\
   $L_{*}$ [L$_{\sun}$]          & 2.88 $\pm$ 0.08       & this work, 7 \\
   Age [Gyr]                     & 2.15 $\pm$ 0.32       & this work, 7 \\

\hline
\end{tabular}
\end{table}

\begin{figure}
    \begin{center}
      \includegraphics[width=0.475\textwidth]{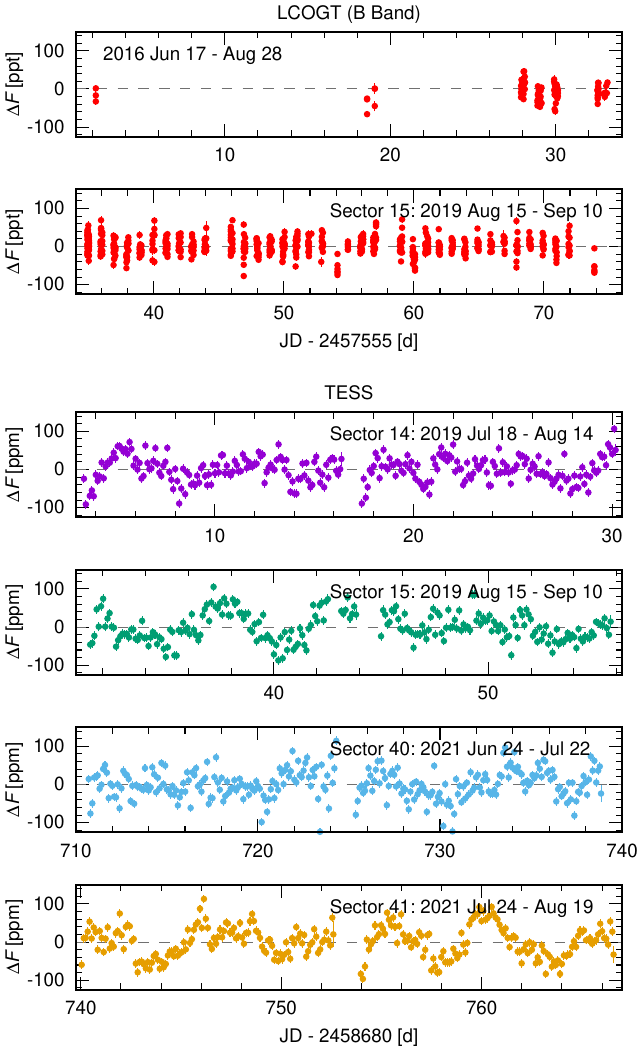}
    \end{center}
    \caption{LCOGT B band lightcurve and {\sc tess} PDCSAP lightcurves for sectors 14, 15, 40 and 41. The data are binned with a 0.1 d interval and are mean-subtracted. The change in flux, $\Delta F$, is plotted in parts per thousand for the LCOGT data and parts per million for the {\sc tess} data.}
    \label{fig:phot_lcs}
\end{figure}

\begin{figure}
    \begin{center}
      \includegraphics[trim=0mm 0mm 0mm 0mm, width=0.47\textwidth]{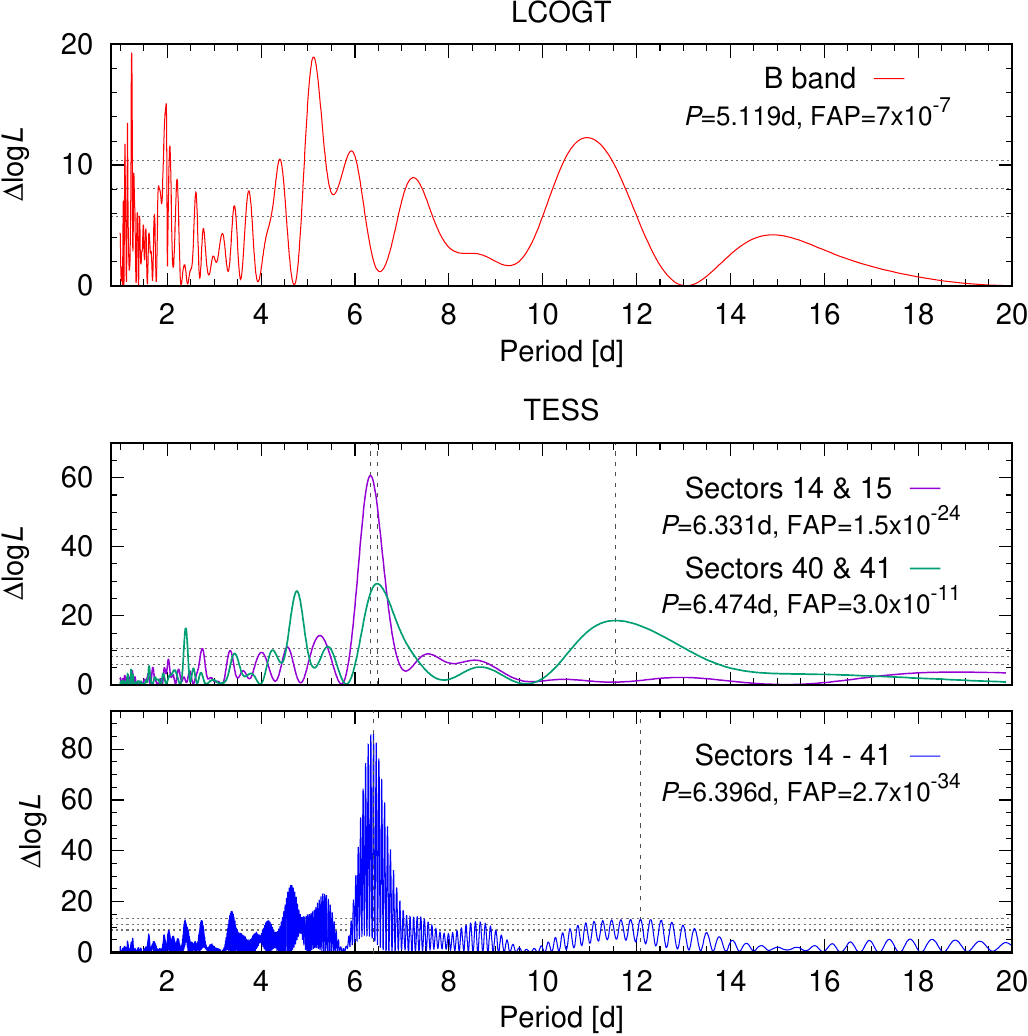}
    \end{center}
    \caption{Generalised Lomb Scargle periodograms.}
    \label{fig:phot_periodogram}
\end{figure}

\section{Photometric observations}
\protect\label{section:photometry}
\subsection{LCOGT photometry}
Photometric observations of \dmppF~were collected using the {Las Cumbres Observatory global telescope network} (LCOGT) robotic telescopes \citep{brown13lcogt}. A total of 761 exposures of $\sim 1$~s duration each were collected over 46 observing nights. Between 3 and 63 observations were made each night, with 1 hour cadence between batches of $3 - 9$ observations. The upper panels of Fig. \ref{fig:phot_lcs} shows the data as mean subtracted flux in parts per thousand (ppt) collected between 2016 June 17 and Aug 28. The most significant peak in the corresponding periodogram (Fig. \ref{fig:phot_periodogram} upper panel) is found at $5.119$~d, with an additional short period peak at $1.979$~d and a longer $10.95$~d peak. The peak at $1.245$~d is the 1-day alias of the $5.119$~d peak. {The peak at $10.95$~d corresponds to a sinusoidal signal with amplitude $5640$~ppm. When phase-folded on this periodicity, the data are found to be clustered into 11 distinct groups.} The $5.119$~d peak corresponds to a sinusoidal signal with semi-amplitude of 7.56 ppt (7560 ppm). We modelled a B-band lightcurve using the image code DoTS \citep{cameron01mapping} for a {\em single} spot with $T_{\rm phot} - T_{\rm spot} = 2000$~K \citep{berdyugina05starspots}, finding that a $r_{\rm spot} = 5.04^{\rm o}$ is required to reproduce this semi-amplitude. {A small spot group with the same effective area as a single spot could also potentially yield the same photometric amplitude, though a more widely distributed collection of spots would likely need to cover a larger total effective area to yield the same sinusoidal or near sinusoidal photometric amplitude. The implied effective spot area} appears to be at odds with the low log($R^{\prime}_{\rm HK}$) activity of \dmppF, especially given that the largest sunspots are never this large when the solar maximum log($R^{\prime}_{\rm HK}$) is still considerably greater than is seen on \dmppF.  Moreover, we find that a $5.04^{\rm o}$ spot would induce a $K = 82$\,ms$^{-1}$ RV signal, which is more than an order of magnitude greater than we find in our RV data (\S \ref{section:spectroscopic}).
We therefore conclude that the periodicities recovered from the LCOGT B band data are not genuine activity-related signals.

\begin{figure*}
    \begin{center}
      \includegraphics[trim=0mm 0mm 0mm 0mm, width=0.99\textwidth]{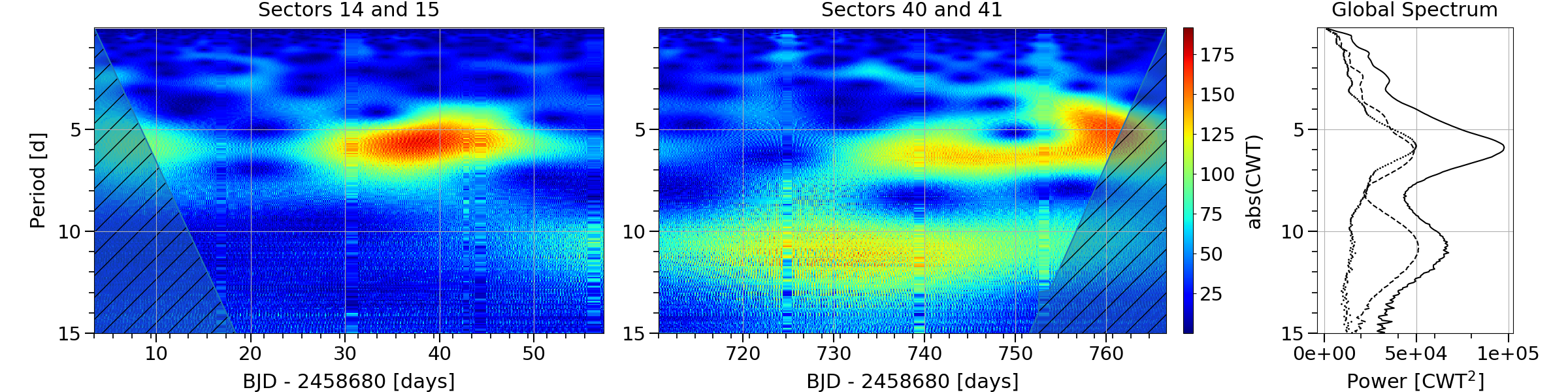}
    \end{center}
    \caption{Continuous wavelet transform (CWT) for the {\sc tess} photometry. The most significant continuous wavelet transform (CWT) powers are shown in red. The shaded area denotes the wavelet cone of influence region. The right hand panel shows the global spectrum for Sectors 14 and 15 (dotted line), Sectors 40 and 41 (dashed line) and all sectors (solid line).}
    \label{fig:phot_wavelet}
\end{figure*}

\subsection{TESS photometry}
\protect\label{section:tess}
The Transiting Exoplanet Survey Satellite ({\sc tess}, \citealt{ricker15tess}) monitored \dmppF~in Sectors 14, 15, 40 and 41. The {\em Mikulski Archive for Space Telescopes}\footnote{https://mast.stsci.edu} provides detrended Pre-search Data Conditioning Simple Aperture Photometry (PDCSAP) lightcurves. Release notes for each {\sc tess} sector\footnote{https://heasarc.gsfc.nasa.gov/docs/tess/documentation.html} (figure 6 of \citealt{fausnaugh18}) give the Combined Differential Photometric Precision (CDPP) for 1hr timescales on $\sim 20000$ targets. The {\sc tess} magnitude for \dmppF~is 5.22 (Table \ref{tab:HD184960params}), indicating an expected CDPP of $\sim 15$~ppm. This scales to $\sim 10$~ppm for the $2.4$~hr ($0.1$~d) bins we used, and is in close agreement with the formal errors on our binned data of $11.4-11.6$~ppm.

Photometric semi-amplitude variability on a scale of \hbox{$< 100$~ppm} (Fig. \ref{fig:phot_lcs}) can be seen at times in the {\sc tess} observations; considerably less than the 7560 ppm seen in the LCOGT data. However, periodic variability is only easily discerned during part of Sectors 15 and 41.
Log-likelihood periodogram searches using sinusoidal signals, are shown in Fig. \ref{fig:phot_periodogram} for Sectors 14 and 15, Sector 40 and 41 and all sectors combined. Highly significant peaks appear in the periodograms with the most significant periods at respectively $6.33$~d, $6.47$~d, $6.40$~d. The respective half width at half maxima (HWHM) of these peaks of $0.33$~d, $0.44$~d and $0.33$~d give an indication of the spread of power, but should not be treated as period uncertainties, as discussed by \cite{vanderplas18}. We also note that the longer period at $11.6$~d (HWHM = $1.4$~d) is also significant in the Sector 40 and 41 periodogram, along with a shorter 4.770 d peak. As with the $6.40$~d strongest peak when using all sectors, the longer peak at $12.1$~d is strongly aliased due to sampling \citep{vanderplas18}. The envelope of the aliased peak possesses HWHM = $1.7$~d. The $12.1$~d peak is marginally significant with $\Delta$log$L = 13.0$ (FAP $= 0.003$).

To investigate the periodicities further, we performed a wavelet analysis using the {\sc scaleogram} package\footnote{https://github.com/alsauve/scaleogram}, which is based on the PyWavelets library \citep{pywavelets}. The resulting continuous wavelet transform in Fig. \ref{fig:phot_wavelet} confirms the visual inspection of periodicities in Fig. \ref{fig:phot_lcs}. The periodicities identified in the log likelihood periodograms are not present throughout the extent of the timeseries and the periodicities are often localised. The global spectrum reveals peak power in Sectors 14 and 15 at $5.8$~d and in Sectors 40 and 41 at $6.0$~d and $10.7$~d (peaking at 11.1~d for the maximum power).

We again used DoTS to model the photometric variability expected from {\sc tess} observations and the absorption line profiles for a cool spot on \dmppF~observed with $550$\,nm central wavelength. The {\sc tess} flux half-amplitude of $\sim 29$\,ppm found by phasing Sectors 15 and 41 is commensurate with an equatorial {\em cool} starspot of radius $0.34^{\rm o}$. A spot of this size is expected to induce RV variability of $K <$\hbox{ $16$\,cms$^{-1}$} at 550\,nm and is thus below the level that could easily be detected by an RV instrument achieving $\sim 1$\,ms$^{-1}$ precision. If a Solar-like facular/spot area ratio of 11.6 is present on \dmppF~(\citealt{shapiro14}; see also \citealt{barnes23moments}), a $0.082^{\rm o}$ region of plage surrounding a $0.024^{\rm o}$ spot could induce a flux half-amplitude of $\sim 29$\,ppm. In this case, the expected RV amplitude is \hbox{$K \sim 38$ cms$^{-1}$} at $550$\,nm. The RV amplitude rises by an order of magnitude for a cool spot with radius $0.2^{\rm o}$ and surrounding plage of $0.68^{\rm o}$. Although the {\sc tess} observations and our spot estimates provide further evidence in support of the low activity of \dmppF, we cannot rule out the possibility of higher activity levels at other epochs when our RV observations were made.

\begin{figure}
    \centering
    \includegraphics[width=0.45\textwidth]{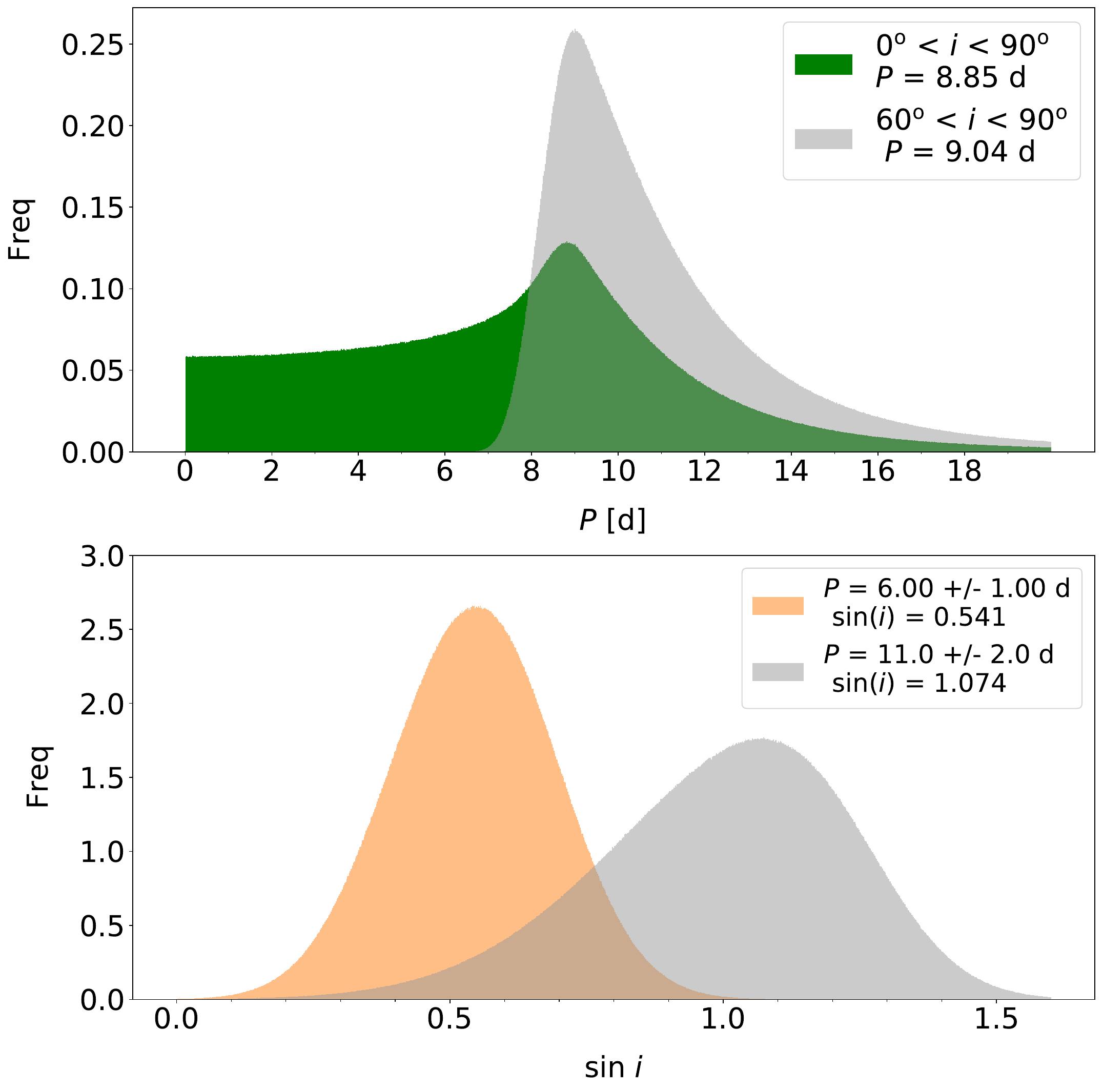}
    \caption{Monte Carlo simulations for stellar rotation period, $P$, and axial inclination, $i$ Top: The distribution of stellar rotation periods based on the parameters in Table \ref{tab:HD184960params}, assuming uniformly distributed periods with the indicated ranged. Bottom: Distribution of stellar axial inclination assuming {\sc tess} periodicities identified in Fig. \ref{fig:phot_periodogram}.}
    \label{fig:montecarlo}
\end{figure}

\subsection{Spot evolution and the orbital period of DMPP-4}
\protect\label{section:spot_period}

A study of sunspot group data spanning over a century by \cite{berdyugina05starspots} found that two active longitudes persist and are separated by $\sim180^\textrm{o}$. One region usually dominates, periodically alternating between the two. More recent work by \cite{basri18doubledipping} has looked at the phenomenon of ``double-dipping'', where a stellar photometric lightcurve spends some of time exhibiting a single sinusoid behaviour at the rotation period and some time in a double-dip mode with two dips or sinusoids within one rotation. The phenomenon was found in the analysis of over 34,000 Kepler lightcurves by \cite{mcquillan14kepler}, where periodogram peaks at the rotation period and half the rotation period were identified. While two distinct active groups {\em may} be present, any spot distribution will typically yield a photometric lightcurve with either one to two dips. For more active stars, spots may thus be distributed over a range of longitudes rather than in two distinct longitude regions. The exact lightcurve morphology is determined by either the relative strength, the distribution of the spots, or by both effects. \cite{basri18doubledipping} used the Kepler lightcurves to examine how much time stars spend in single-dip mode and how much in double-dip mode. For their hottest sample corresponding to stars with $T_{\rm eff} = 6000-6200K$, the time spent in double-dip mode is 1.8 times that spent in single-dip mode. This bimodal behaviour is seen in the {\sc tess} observations of \dmppF. At the {\sc tess} Sector 14 and 15 epochs, two spot groups of similar intensity could yield the periodic signatures seen at $\sim 5.5-6.5$~d, while in Sectors 40 and 41, there appears to be evidence for one spot group for some of the time {\em and} two spot groups for some of the time. Evolution of spot groups appears to be relatively rapid and on the timescale of a single rotation, resulting in a rapidly changing lightcurve. The lightcurve in Sectors 40 and 41 can be explained by a single weak/small spot that grows in strength with the simultaneous appearance of a second spot group resulting in a switch from single-dip to double-dip mode.

In summary, we believe that \dmppF~exhibits small spots, as evidenced by the low log($R^{\prime}_{\rm HK}$), photometric amplitude and spot size estimate. It possibly follows the solar paradigm of two persistent and distinct active regions rather than that of potentially more distributed activity on an active star. However, interpreting periodicities at $P/2$ from this scenario may be simplistic since other effects such as differential rotation and spot evolution can result in both double dipping and multiple periodicities.
A rotation period of $\sim 11$~d rather than $\sim 6$~d thus seems more likely given the estimated stellar rotation {velocity} and radius.

\subsection{Period and inclination distributions}
\protect\label{section:montecarlo}
We expect DMPP to find systems with high orbital inclinations where the stellar Ca~{\sc ii}~H\&K emission would be more readily obscured \citep{haswell12wasp12,haswell20dmpp}. So,
under the assumption that the stellar spin axis and planetary orbital planes are aligned, we expect a stellar rotation axis which is highly inclined to our line of sight.  The upper plot in Fig. \ref{fig:montecarlo} shows the distribution of expected periods from the $R_* = 1.38 \pm 0.01$~R$_\odot$ (Gaussian distribution) and $v$\,sini\,$i$ = $7.5^{+0.5}_{-2.5}$\,ms$^{-1}$ (asymmetric Gaussian distribution) estimates listed in Table \ref{tab:HD184960params}. A uniform distribution of axial inclinations of $0^{\rm o} < i < 90^{\rm o}$ was simulated. We find a resulting modal period of $P$ = $8.89$~d with a range between $2.74$~d~to $11.10$~d at the 16 per cent and 84 percent intervals. A simulation with $60^{\rm o} < i < 90^{\rm o}$ only affects the most likely period slightly ($P$ = $9.02$~d), but skews the posterior distribution to higher values of $P$ ($8.69$~d to $13.50$~d). The lower panel in Fig. \ref{fig:montecarlo} shows the axial inclination distributions, $i$, assuming period distributions with $P = 6.0$~d and $11.0$~d from the TESS periodicities identified in \S \ref{section:tess} (assuming Gaussian distributions of $\sigma = \pm1$~d and $\pm2$~d). For the $P = 6.0$~d case, a low axial inclination with modal value, $i = 32.8^{\rm o}$ ($23.2^{\rm o}$ to $43.8^{\rm o}$ at 16 per cent and 84 percent intervals) is found. For $P = 11.0$~d, inclinations $>49.7^{\rm o}$ (16 per cent) are found with a modal sin~$i > 1$. The large uncertainty in our \vsini~estimate, particularly in the lower limit is a likely explanation for the difference between the $11$~d photometric rotation period and Monte Carlo modal period of $8.89$~d. If $i=90^{o}$, \vsini~must be $\leq 6.35$\,kms$^{-1}$ when $P = 11$~d.

\begin{table*}
	\centering
  \caption[\dmppF~activity correlations]{\dmppF~correlations between RV and chromatic index, $\kappa$, full width at half maximum (FWHM), RV vs bisector span (BIS) and RV vs Ca H\&K S-index. Pearson's r, and student's $p$ statistic are shown for each correlation for all SOPHIE observations combined and all HARPS-N observations combined (top rows). The remaining rows show the same statistics for individual observing runs.}
  \protect\label{tab:correlations}

  \begin{tabular}{lcccc}

\hline
           & Kappa                & FWHM                 & BIS                  & S-index  \\
           & $r$ ($p$)            & $r$ ($p$)            & $r$ ($p$)            & $r$ ($p$) \\
\hline
All SOPHIE &   0.019  (  0.902  ) &    0.057  (  0.712  ) &  -0.366  (  0.014  ) &   -0.054  (  0.726  ) \\
All HARPS-N &  0.287  (  0.146  ) &   -0.204  (  0.308  ) &  -0.255  (  0.199  ) &    0.236  (  0.236  ) \\
\hline
SO-15A  &   0.178  (  0.703  ) &    0.299  (  0.514  ) &   0.544  (  0.207  ) &   -0.119  (  0.799  ) \\
SO-15B  &   0.032  (  0.925  ) &    0.253  (  0.452  ) &  -0.406  (  0.215  ) &    0.028  (  0.934  ) \\
SO-16A  &  -0.208  (  0.496  ) &   -0.268  (  0.376  ) &  -0.601  (  0.030  ) &   -0.114  (  0.712  ) \\
SO-16B  &  -0.271  (  0.350  ) &   -0.613  (  0.020  ) &  -0.732  (  0.003  ) &   -0.242  (  0.405  ) \\
HN-16A  &   0.405  (  0.096  ) &   -0.235  (  0.348  ) &  -0.317  (  0.201  ) &    0.403  (  0.097  ) \\
HN-19A  &  -0.007  (  0.986  ) &   -0.287  (  0.454  ) &   0.540  (  0.134  ) &   -0.018  (  0.962  ) \\

\hline

\end{tabular}
\end{table*}

\begin{figure*}
\begin{center}
\begin{tabular}{cc}
	\includegraphics[trim=0 1mm 0 0, height=1.49\columnwidth,width=1.00\columnwidth]{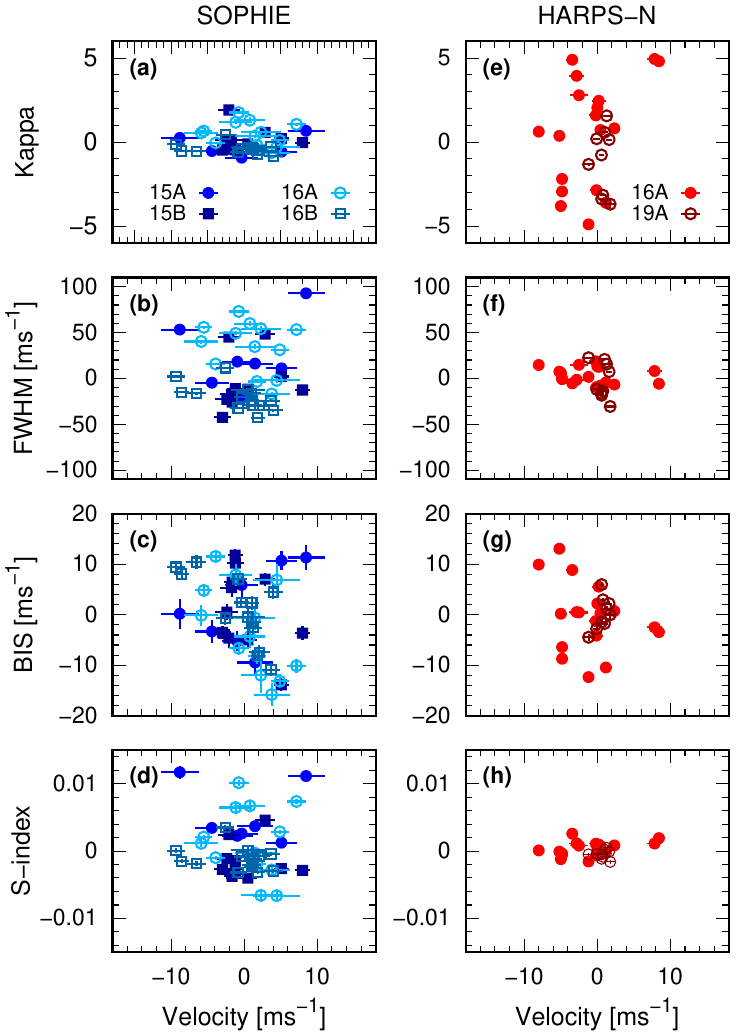}
    \hspace{2mm} &
    \includegraphics[width=1.0\columnwidth]{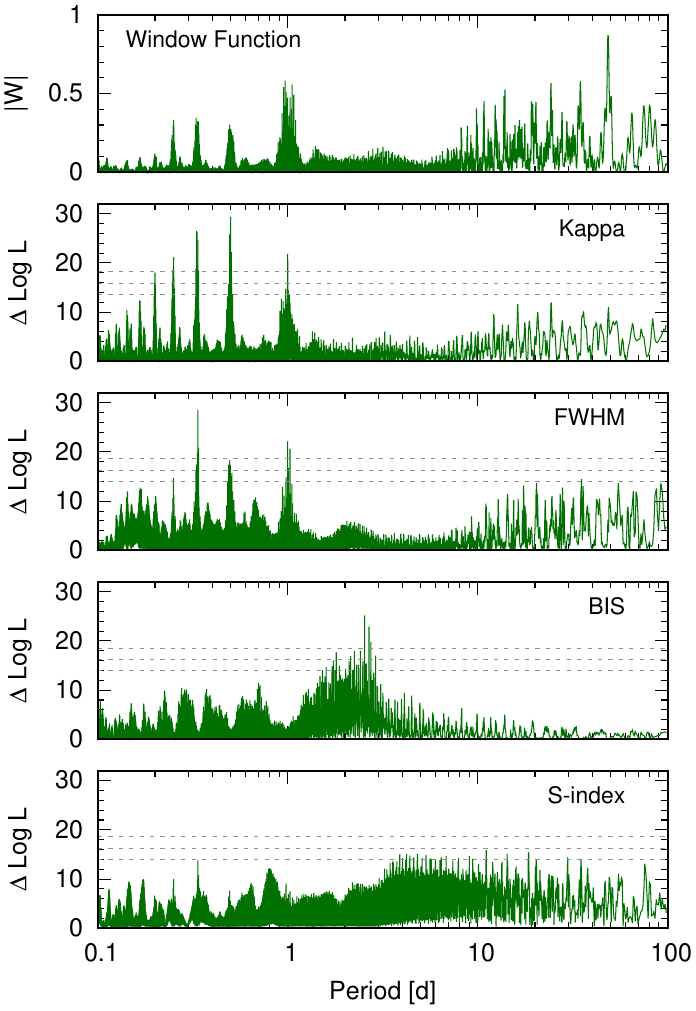}
    \\
    \end{tabular}
    \end{center}
    \caption{Left: Activity index correlations with RV for SOPHIE (panels a-c) and HARPS-N (panels d-f) observations. The points are colour coded according to the observation semester. Right: Activity index periodograms for line full width at half maximum (FWHM), bisector inverse span (BIS) and Calcium S-index. FAP at 0.1\%, 1\% and 10\% are (top to bottom) are indicated by the dashed horizontal lines. }
    \label{fig:correlations}
\end{figure*}

\begin{figure}
    \centering
    \includegraphics[width=0.4\textwidth]{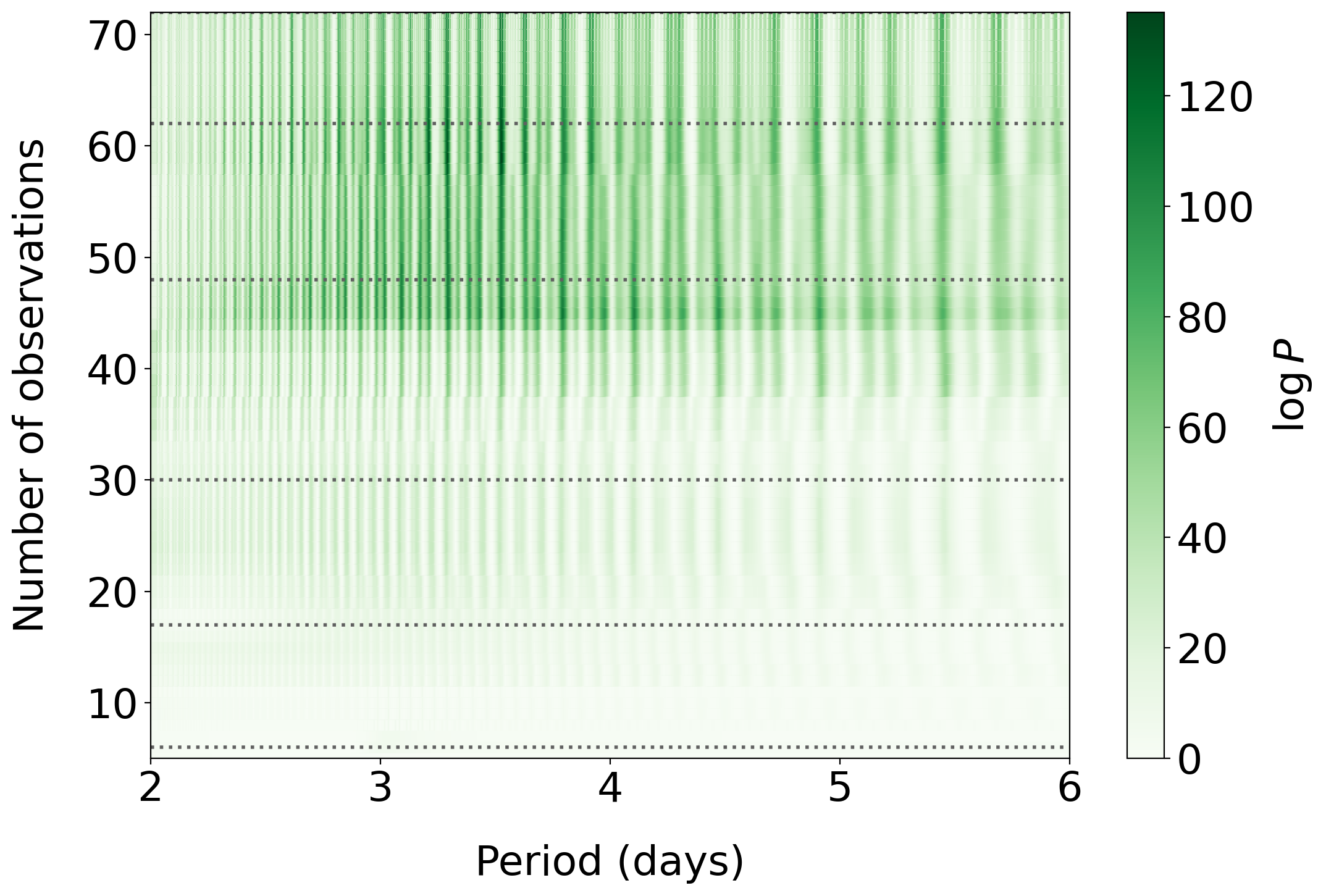}
    \includegraphics[width=0.4\textwidth]{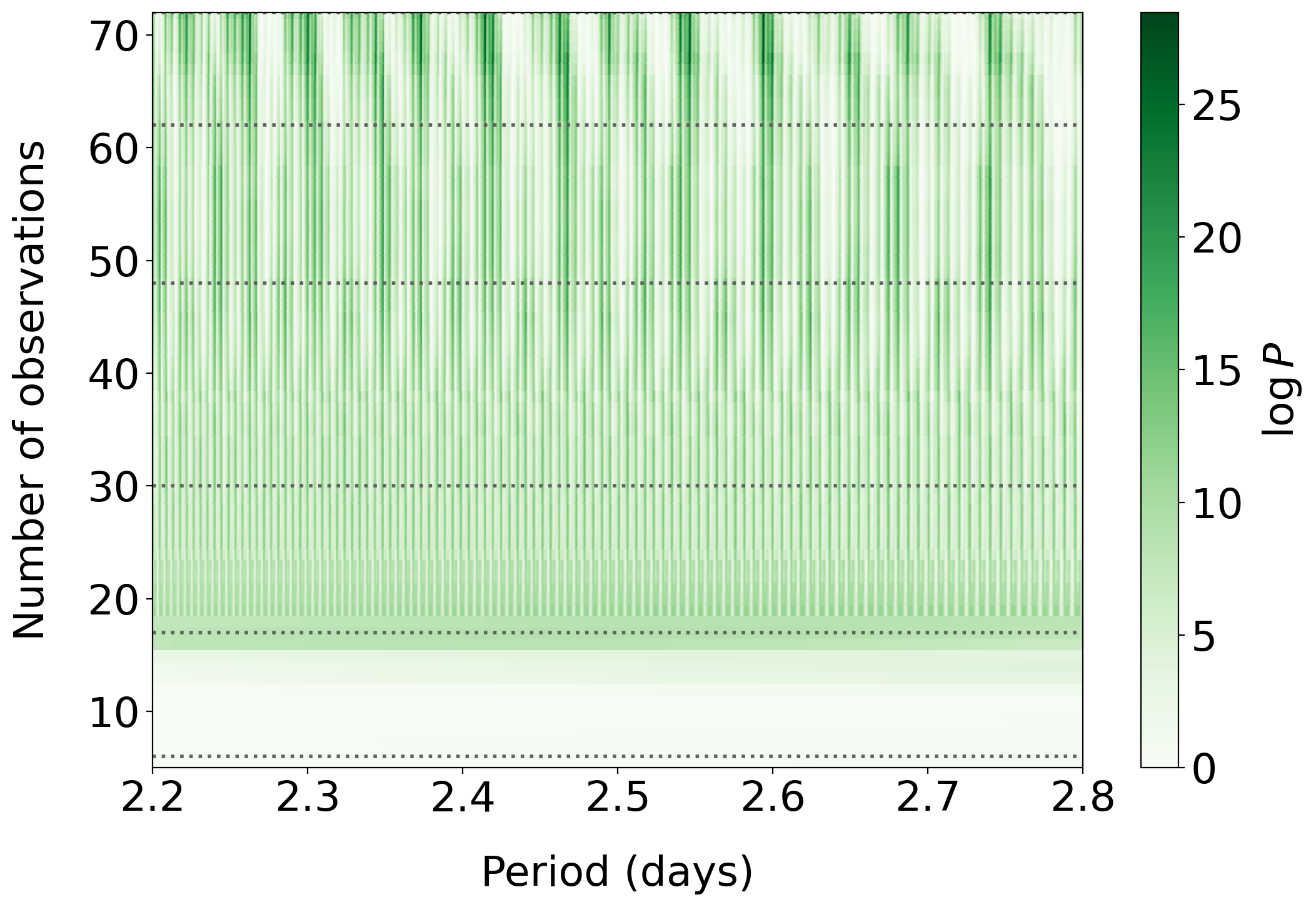}
    \caption{Stacked periodograms of the RVs (top) and residual RVs after accounting for the first signal (bottom). The dotted lines indicate the end of each observing run.}
    \label{fig:stacked}
\end{figure}

\begin{figure*}
	\begin{center}
	\includegraphics[trim=0 4mm 0 0, width=1.90\columnwidth]{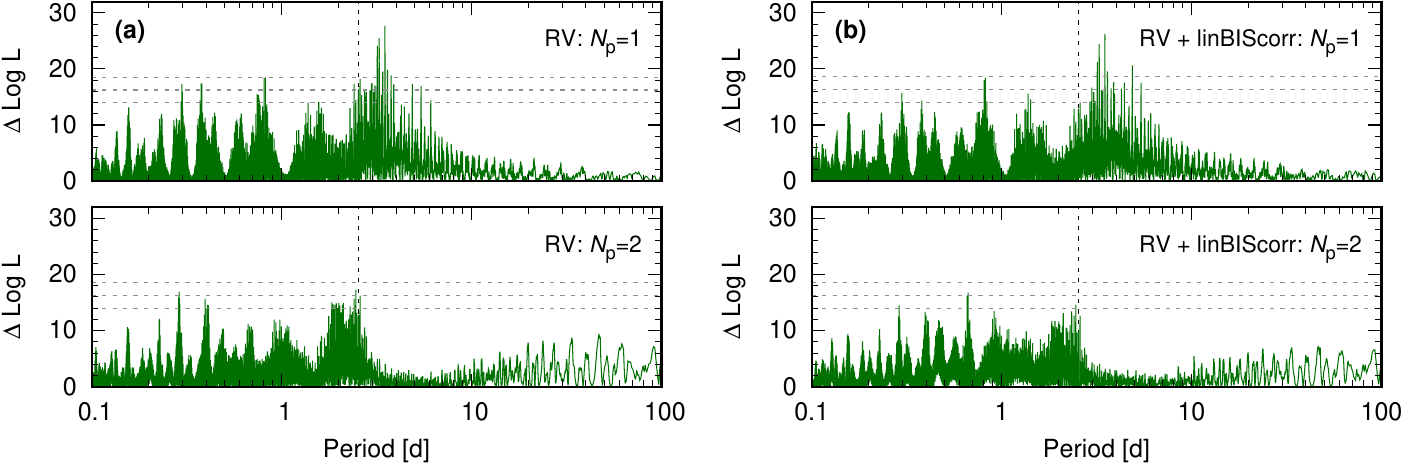}
	\end{center}
    \caption{Radial velocity log likelihood periodograms using (a) RVs only and (b) with a BIS linear correlation (RV+linBIScorr). The periodograms for single ($N_{\textrm{p}} = 1$) Keplerian signals are shown in the top panels. The period searches with a second recursively added signal ($N_{\textrm{p}} = 2$) are shown in the bottom panels. The vertical dashed line indicates the BIS period of 2.54~d. Horizontal dashed lines are the 0.1\%, 1\% and 10\% FAPs as in Fig. \ref{fig:correlations}.}
    \label{fig:rv_periodograms}
\end{figure*}


\section{Activity indicators}
\protect\label{section:spectroscopic}

Spectroscopic observations of \dmppF~were made with with SOPHIE \citep{perruchot08sophie} at the 1.9\,m Observatoire de Haute Provence Telescope during four semesters in 2015 and 2016. Further data were obtained with HARPS-N \citep{cosentino14tng} at the Telescopio Nazionale Galileo telescope. {Observations were made in 2015 Apr-May on 3 nights over a time span of 4 nights; and again later in \hbox{2015 Nov} on 5 nights. Similarly, observing runs in \hbox{2016 May} and \hbox{2016 Nov-Dec} each resulted in 4 nights of observations over a 5 night time span. We refer respectively to the individual SOPHIE runs as \hbox{SO-15A}, \hbox{SO-15B}, \hbox{SO-16A} and \hbox{SO-16B}. In 2016, HARPS-N observations on 4 nights \hbox{(HN-16A)} were made in \hbox{2016 Jul} between the \hbox{SO-16A} and \hbox{SO-16B} runs. The last set of HARPS-N observations were made in \hbox{2019 Aug} on a single night \hbox{(HN-19A)}.} Exposures were monitored to ensure relatively stable counts with SOPHIE. Because \dmppF~is very bright, observation times of $200$\,s to $900$\,s were used, depending on observing conditions. When observations were cycled with other targets, exposures on \dmppF~were made for at least $\sim900$\,s in a single block to ensure that possible 5 minute stellar oscillations were minimised.

The RVs were {derived using} the matched-template code, {\sc harps-terra} \citep{anglada12harpsterra}. We also used the activity indices provided by the standard data reduction software pipelines for both instruments. A total of 111 observations were made with SOPHIE, while 87 observations were made with HARPS-N. After {optimally weighted binning into $900$\,s observations, the respective data sets comprise} 45 and 27 observations. The binned data are tabulated in Tables \ref{tab:sophie_rvs_and_activity} \& \ref{tab:harps_rvs_and_activity}.

The low altitude of the Observatoire de Haute Provence, small telescope aperture and instrumental effects such as reference/target fibre drift, potential chromatic $\kappa$-correlation (see below) and charge transfer inefficiency effects limit the precision that is achieved with SOPHIE. The formal uncertainties for our 2015 and 2016 observations, {before $900$\,s binning, are $2.25$\,ms$^{-1}$ and $2.26$\,ms$^{-1}$. The binned data uncertainties are respectively $1.69$\,ms$^{-1}$ and $1.60$\,ms$^{-1}$. A measurement precision of $\sim 1-2$\,ms$^{-1}$ has been demonstrated with SOPHIE on RV standard stars on timescales of a few tens of days \citep{bouchy13sophie}. For HARPS-N, the formal uncertainties for single $200$~s observations is $1.17$\,ms$^{-1}$, and $0.73$\,ms$^{-1}$ for $900$~s binned data.}


\subsection{Kappa correlation with RVs}

We investigated correlations between possible diurnal chromatic systematics and the RVs. \cite{bourrier14} and \cite{berdinas16} identified a significant intra-night systematic effect on observations made with the HARPS-N spectrograph. The primary cause appears to be an incomplete correction of differential atmospheric refraction by the atmospheric dispersion corrector (ADC), causing colour-dependent (chromatic) flux-losses. The effect is thus potentially important for targets observed throughout the night at different airmasses. We measured the `chromatic index', $\kappa$, defined as the slope of a linear portion of the pseudo spectral energy distribution about a specified wavelength. In practice, we measured the slope in mean flux per \'{e}chelle order across several orders in a similar manner to \cite{berdinas16}; see \cite{haswell20dmpp} for full details. Fig.~\ref{fig:correlations} (left panel) shows the chromatic index, $\kappa$, plotted against the RVs for all SOPHIE and HARPS-N observations. The degree of correlation is measured via Pearson's $r$ coefficient and listed in Table \ref{tab:correlations}; only weak or very weak correlations are seen during most observing runs. There is a moderate correlation in the HN-16A data set, but the significance is also moderate, as indicated by the student's t test p-value. This suggests that there is no clear evidence of a correlation. The correlation is driven by the two outlying RVs (greatest positive values); removing them yields $r=0.05$ and $p=0.86$.

\subsection{RV correlations with activity measures}

The correlation of RVs with the full width at half-maxiumum (FWHM), line Bisector Inverse Span (BIS) and Ca~{\sc ii}~H\&K S-index are also plotted in Fig. \ref{fig:correlations}. Correlation measurements are listed in Table \ref{tab:correlations}. We include the HN-19A data in Table \ref{tab:correlations} for completeness; since this run comprised a single night, activity correlations are likely to be less meaningful.

The BIS shows significant moderate anti-correlation in the SO-16A and SO-16B data sets, but is less pronounced in SO-15B.  We note that the SO-15A data is of significantly lower precision than the subsequent observations. For BIS in SO-15A, removing the data point with largest positive RV~$= 8.49$\,ms$^{-1}$ results in $r=-0.64$ and $p=0.17$. For individual observing runs, the BIS anti-correlation is largest and most significant in the SO-16A and SO-16B epochs. In SO-16B, the RVs also show a significant (\textit{p}=0.02) moderate anti-correlation with FWHM.

HARPS-N shows moderate correlations, but the significance is low. The HN-16A run shows a moderate S-index variability, although the probability of no correlation is still nearly 10\% (i.e. $p = 0.097$). Of the three activity indicators, only BIS shows a consistently moderate-good correlation.


\begin{table*}
\centering
\caption[\dmppF~likelihoodperiods]{Candidate maximum likelihood periodogram search summary for (a) RV period search and (b) RV period search with a linear BIS correlation (RV+linBIScorr model). The corresponding Bayesian Information Criteria (BIC) are also given along with the $\Delta\textrm{BIC}$ values for $\textrm{BIC}(N_\textrm{p} - (N_\textrm{p}-1))$ in parentheses and directly derived estimates of the corresponding Bayes Factors (BF).}
\protect\label{tab:likelihood}
\begin{tabular}{cccccccccc}
\hline
{Keplerians} &       & \multicolumn{3}{c}{(a) RV}                                                         & ~~ &       & \multicolumn{3}{c}{(b) RV+linBIScorr}                                     \\
\hline
$N_\textrm{p}$     & P [d] & logL ($\Delta\mathrm{log}L$)  & BIC ($\Delta\mathrm{BIC}$)  & BF                &    & P [d] &  logL ($\Delta\mathrm{log}L$) & ($\Delta\mathrm{BIC}$)    & BF                   \\
\hline
0          &    -  &   -200.2                      & 417.6                      &                    &    & -     & -195.0                       & 419.7                      &                      \\
1          & 3.498 &   -172.6 (27.7)               & 375.1 (-42.5)              & $1.7\times10^{9}$ &    & 3.498 & -168.7 (26.3)                & 382.1 (-37.8)              & $1.6 \times 10^{8}$  \\
2          & 2.459 &   -155.2 (17.3)               & 353.2 (-21.9)              & 56954             &    & 2.459 & -154.1 (14.6)                & 367.8 (-14.3)              & 1274                 \\
\hline
\end{tabular}
\end{table*}

\subsection{Periodicities in activity measures}
\protect\label{section:activityperiods}
The right panels in Fig. \ref{fig:correlations} show periodograms for the $\kappa$-correlation and each activity index. Aliases and integer fractions of a day in the Kappa and FWHM periodograms are seen. The $\kappa$-correlation and FWHM show no other significant peaks. For BIS, we find a 2.54 d peak with FAP = $2.7 \times 10^{-6}$ and $\Delta$~Log$L = 27.2$ relative to the scenario with no signal. A neighbouring 2.68 d peak with log$L_{2.54-2.68} <5$ ($\Delta$~Log$L = 22.9$) relative to the main peak is also present. This periodicity, taken with the moderate anti-correlation between BIS and RV, suggests that the RVs may be affected by stellar activity. It is nevertheless difficult to reconcile this finding with our analysis that reveals very low photometric variability and the low log($R'_{\rm HK}$), suggesting that \dmppF~is an inactive star. The variable degree of correlation between BIS and RV at different observing epochs suggests that the star shows variable activity. This should however be treated with some caution since each observing epoch comprises only a few nights of observations and so may not sample a complete stellar rotation.

{Further, the most significant BIS periodicity is likely to be a fraction of the true period. Period analysis of apparent RV shifts due to line-shape changes as described by BIS, are a proxy measure of the third central moment or line skewness. \cite{barnes23moments} finds that for stars with realistic spot and facular distributions and solar activity levels, RV periodicities may appear with dominant power at harmonics of the rotation period, including $P_\textrm{rot}/2$ and $P_\textrm{rot}/4$. This behaviour was first noted by \citet{boisse11}. For instance, a single high latitude spot can easily induce RV and BIS variability at $P_\textrm{rot}/2$, while a low latitude spot could simultaneously induce RV variability at $P_{\rm rot}/2$ and BIS variability at $P_\textrm{rot}/4$. A pair of low-latitude spots located at active longitudes 180$^\textrm{o}$ apart, as has been observed for the sun \citep{berdyugina03activelongs}, could then also be expected to lead to predominant periodicities in RV and BIS at $P_\textrm{rot}/4$. The exact periodicities are likely to vary from star to star, but depend on the exact spot patterns, presence of absence of faculae and the combination of spot latitude and stellar axial inclination and $v\,\textrm{sin}\,i$.}

The BIS periodicities of $2.54$~d and $2.68$~d may thus be manifestations of active regions that have true periodicities at twice or four times these values. In other words, $P_{\rm rot}$ inferred from BIS may be either $5.08$~d and $5.36$~d ($2\times$) or $10.2$~d and \hbox{$10.7$~d (4$\times$)}. The longer periods are close to photometric periodicities identified from the {\sc tess} Sector 40 and 41 observations in \S \ref{section:tess}. There is tentative evidence for low significance peaks (1\,-\,10 per cent FAP) at $\geq 10$~d in the S-index periodogram, which shows a most significant peak at $11.1$~d ($\Delta$~Log$L = 15.9$), again in close agreement with the {\sc tess} observations.


\begin{table*}
    \begin{tabular}{p{4cm}cccc}
    \hline
    {Parameter} & \multicolumn{3}{c}{Model fits to RV data only \vspace{2mm}} & {RV + BIS} \\
     & {\em N}$\mathbf{_{p}=1}$ & {\em N}$\mathbf{_{p}=2}$ & {\em N}$\mathbf{_{p}=1}$ {+~GP} & {\em N}$\mathbf{_{p}=1}$ {+~GP} \\
    \hline
    $T_{\rm 0,b}$ [d] & $2457139.7^{+1.5}_{-1.2}$ & $2457140.26^{+0.81}_{-0.65}$ & $2457139.8^{+1.3}_{-1.3}$ & $2457139.9^{+1.2}_{-1.3}$ \\
    $P_{\rm b}$ [d] & $3.49791^{+0.00046}_{-0.00143}$ & $2.4570^{+0.0026}_{-0.0462}$ & $3.4982^{+0.0015}_{-0.0027}$ & $3.4982^{+0.0022}_{-0.2327}$ \\
    $K_{\rm b}$ [ms$^{-1}$] & $4.24^{+0.55}_{-0.56}$ & $3.26^{+0.46}_{-0.57}$ & $4.58^{+0.59}_{-0.67}$ & $4.43^{+0.66}_{-0.69}$ \\
    $e_{\rm b}$ & <0.067 & <0.065 &  <0.063  &  <0.063 \\
    $\omega_{\rm b}$ [rad] & $3.9^{+1.4}_{-2.7}$ & $2.7^{+2.4}_{-1.6}$ & $3.6^{+1.8}_{-2.5}$ & $3.5^{+1.9}_{-2.3}$ \\
    $a_{\rm b}$ [$\rm AU$] & $0.04853^{+0.00028}_{-0.00047}$ & $0.03836^{+0.00039}_{-0.00044}$ & $0.04854^{+0.00033}_{-0.00054}$ & $\hspace{5mm}0.04853^{+0.00036}_{-0.00197}\hspace{5mm}$ \\
    $m_p$\,sin\,$i$\,$_{\rm b}$ [M$_\oplus$] & $11.6^{+1.5}_{-1.6}$ & $8.0^{+1.1}_{-1.5}$ & $12.6^{+1.6}_{-1.8}$ & $12.2^{+1.8}_{-1.9}$ \\
    \hline
    \multicolumn{5}{c}{Planet candidate c\vspace{2mm}} \\
    $T_{\rm 0,c}$ [d] & & $2457138.9^{+1.6}_{-1.8}$ &    &   \\
    $P_{\rm c}$ [d] &   & $5.4196^{+0.6766}_{-0.0030}$ &    &   \\
    $K_{\rm c}$ [ms$^{-1}$] &   & $3.81^{+0.43}_{-0.50}$ &    &   \\
    $e_{\rm c}$ &   & <0.064 &    &   \\
    $\omega_{\rm c}$ [rad] &    & $2.8^{+2.2}_{-1.7}$ &    &   \\
    $a_{\rm c}$ [$\rm AU$] &  & $0.06514^{+0.00469}_{-0.00041}$ &  &  \\
    $m_p$\,sin\,$i$\,$_{\rm c}$ [M$_\oplus$]  &   & $12.2^{+1.4}_{-1.6}$ &   &   \\
    \hline
    \multicolumn{5}{c}{Quasi Periodic GP parameters\vspace{2mm}} \\
    $\eta_1$ &  &   & $2.4^{+1.3}_{-1.8}$ & $2.88^{+1.09}_{-1.03}$ \\
    $\eta_{1,2}$ & &   &   & $-8.3^{+1.8}_{-2.3}$ \\
    $\eta_2$ &   &   & $20.1^{+4.9}_{-5.0}$ & $19.9^{+5.0}_{-4.8}$ \\
    $\eta_3$    &   &   & $2.42^{+0.25}_{-0.24}$ & $2.51^{+0.09}_{-0.10}$ \\
    $\Gamma~(\eta_4=\sqrt{2/\Gamma}$) [fixed] &   &   & $1~(\sqrt{2})$ & $1~(\sqrt{2})$ \\
    \hline
    \multicolumn{5}{c}{Systemic RVs and white noise parameters\vspace{2mm}} \\
    $\gamma_\textrm{RV,HN}$ [ms$^{-1}$] & $-0.93^{+0.55}_{-0.56}$ & $-0.55^{+0.64}_{-0.70}$ & $-0.50^{+1.79}_{-1.57}$ & $0.2^{+1.3}_{-1.2}$ \\
    $\rm{Offset_{RV,SO}}$ [ms$^{-1}$] & $0.63^{+0.75}_{-0.74}$  & $0.37^{+0.97}_{-0.76}$  & $0.54^{+2.06}_{-1.94}$  & $-0.49^{+2.5}_{-2.4}$ \\
    $\sigma_\textrm{RV,SO}$ [ms$^{-1}$] & $2.60^{+0.48}_{-0.45}$  & $0.41^{+0.92}_{-0.37}$  & $1.25^{+0.52}_{-0.64}$  &  $1.50^{+0.40}_{-0.33}$ \\
    $\sigma_\textrm{RV,HN}$ [ms$^{-1}$] & $1.75^{+0.45}_{-0.47}$  & $1.15^{+0.51}_{-0.77}$ & $1.49^{+0.40}_{-0.32}$  & $1.29^{+0.49}_{-0.44}$ \\
    $\sigma_\textrm{BIS,SO}$ [ms$^{-1}$] &   &   &   &  $2.80^{+0.68}_{-0.56}$ \\
    $\sigma_\textrm{BIS,HN}$ [ms$^{-1}$] &   &   &   & $4.69^{+1.05}_{-0.96}$ \\

    \hline
    \multicolumn{5}{c}{Fitting statistics\vspace{2mm}} \\
    BF & >7676 & 2397 & 111.4 & 1038 \\
    log $L$ & -168.4 & -147.9 & -162.8  & -394.6 \\
    \end{tabular}
    \caption{Posterior RV Model parameters and derived planet masses with 68.3\% confidence uncertainties.The log~$L$ statistics in the final row are given for the best fitting (maximum posterior) models. Models with 1-Keplerian ($N_\textrm{p}=1$) and 2-Keplerians ($N_\textrm{p}=2$) are tabulated in columns 1 and 2. Column 3 is the $N_\textrm{p}=1$ + GP solution using only the RVs. Column 4 tabulates parameters for the $N_\textrm{p}=1$ + GP solution which uses simultaneous RVs and BIS. Eccentricities are all consistent with 0 and therefore upper limits are quoted at 2$\sigma$.}
    \label{tab:bigsolution}
\end{table*}


\begin{figure*}
        \begin{subfigure}[b]{0.65\textwidth}
         \caption{}
         \includegraphics[width=\textwidth, trim=0mm 0mm 0mm 0mm]{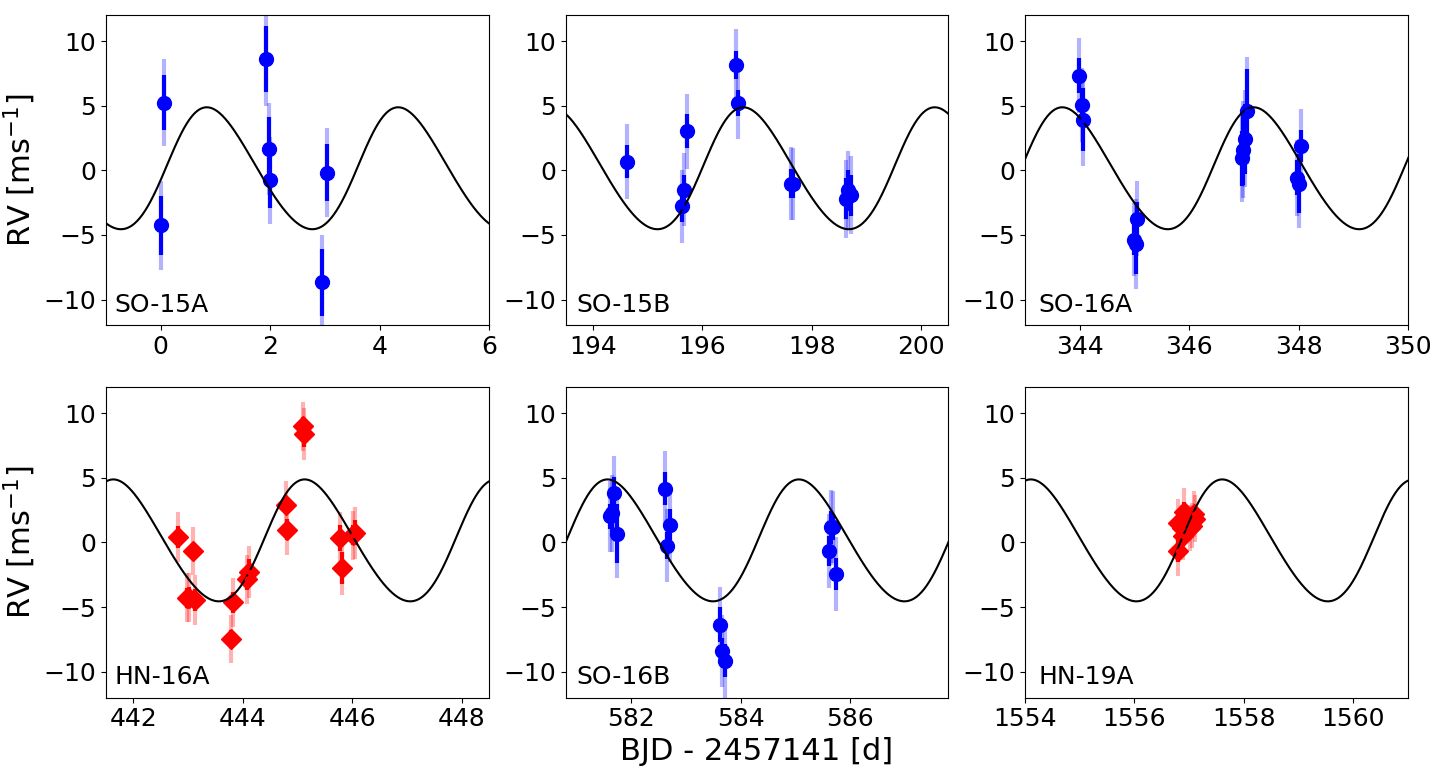}
         \label{fig:1kep_a}
     \end{subfigure}
     \hfill
          \begin{subfigure}[b]{0.325\textwidth}
         \caption{}
         \includegraphics[width=\textwidth, trim=2mm -20mm 2mm 0mm]{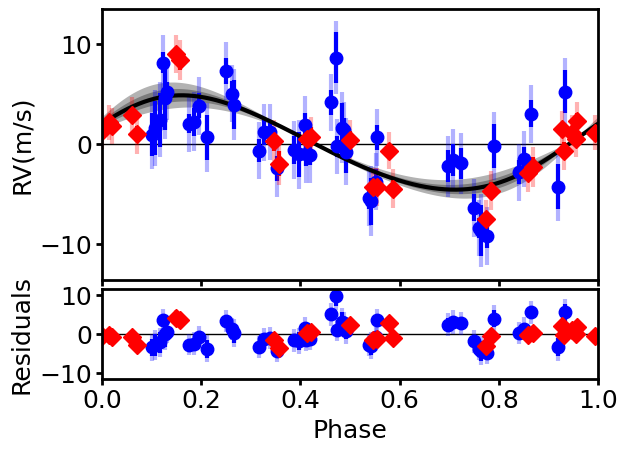}
         \label{fig:1kep_b}
     \end{subfigure}
    \caption{Single Keplerian RV solution ($N_{\textrm{p}}=1$) for combined SOPHIE (SO) and HARPS-N (HN) data sets taken in semesters 15A, 15B, 16A, 16B and 19A. {(a)} RVs vs observation time and {(b)} phased RVs showing fit with 68\% and 95\% model uncertainties. The light blue and pink data uncertainties indicate the combined formal uncertainties and additive white noise.}
    \label{fig:rv_1kep}
\end{figure*}

\begin{figure*}
        \begin{subfigure}[b]{0.65\textwidth}
         \caption{}
               \includegraphics[width=\textwidth, trim=0mm -80mm 0mm 0mm]{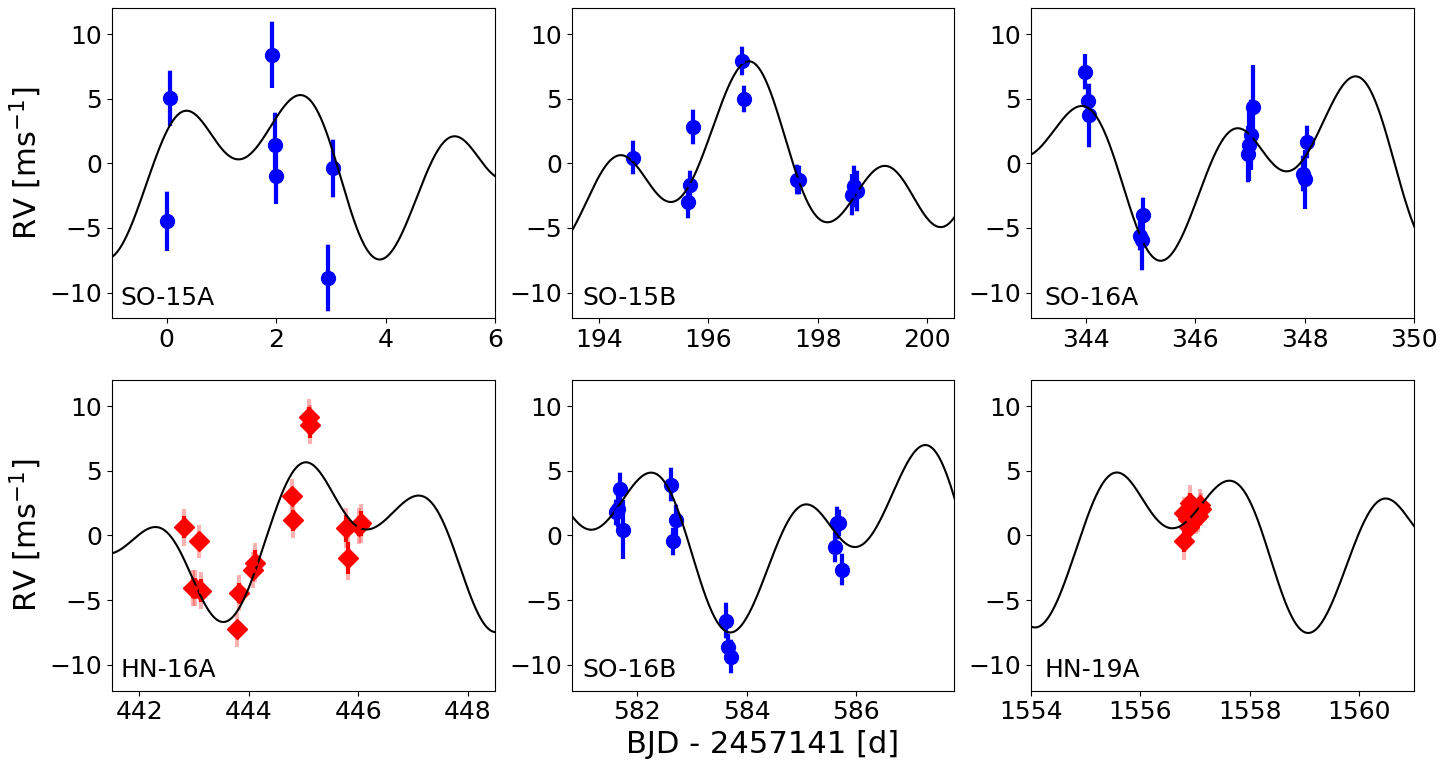}
         \label{fig:2kep_a}
     \end{subfigure}
     \hfill
     \begin{subfigure}[b]{0.325\textwidth}
         \caption{}
      \includegraphics[width=\textwidth, trim=2mm   1mm 2mm 0mm]{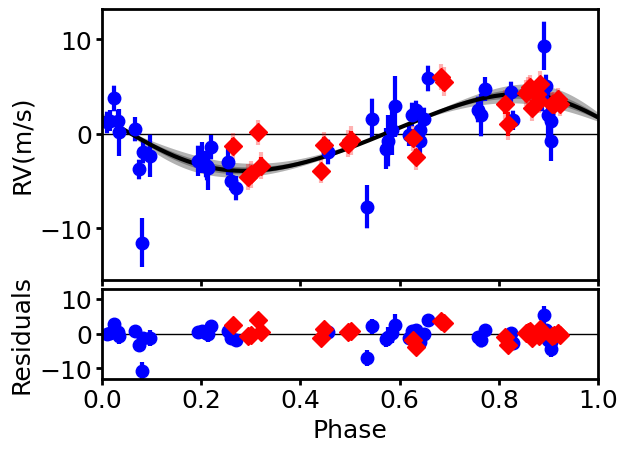} \\
      \includegraphics[width=\textwidth, trim=2mm -20mm 2mm 0mm]{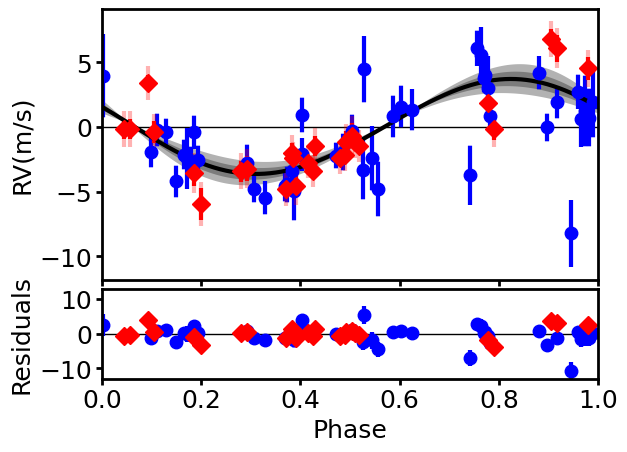} \\

         \label{fig:2kep_b}
     \end{subfigure}
    \caption{As in Figure \ref{fig:rv_1kep} for the 2-Keplerian ($N_{\textrm{p}}=2$) solution. (a) The RVs with solution curve and (b) the phased RVs plotted for the $5.419$~d candidate signal (top) and the $2.457$~d signal (bottom).}
    \label{fig:rv_2kep}
\end{figure*}

\begin{figure*}
    \begin{center}
     \begin{subfigure}[b]{0.65\textwidth}
         \caption{}
         \includegraphics[trim=0mm 0mm 0mm 0mm, width=\textwidth]{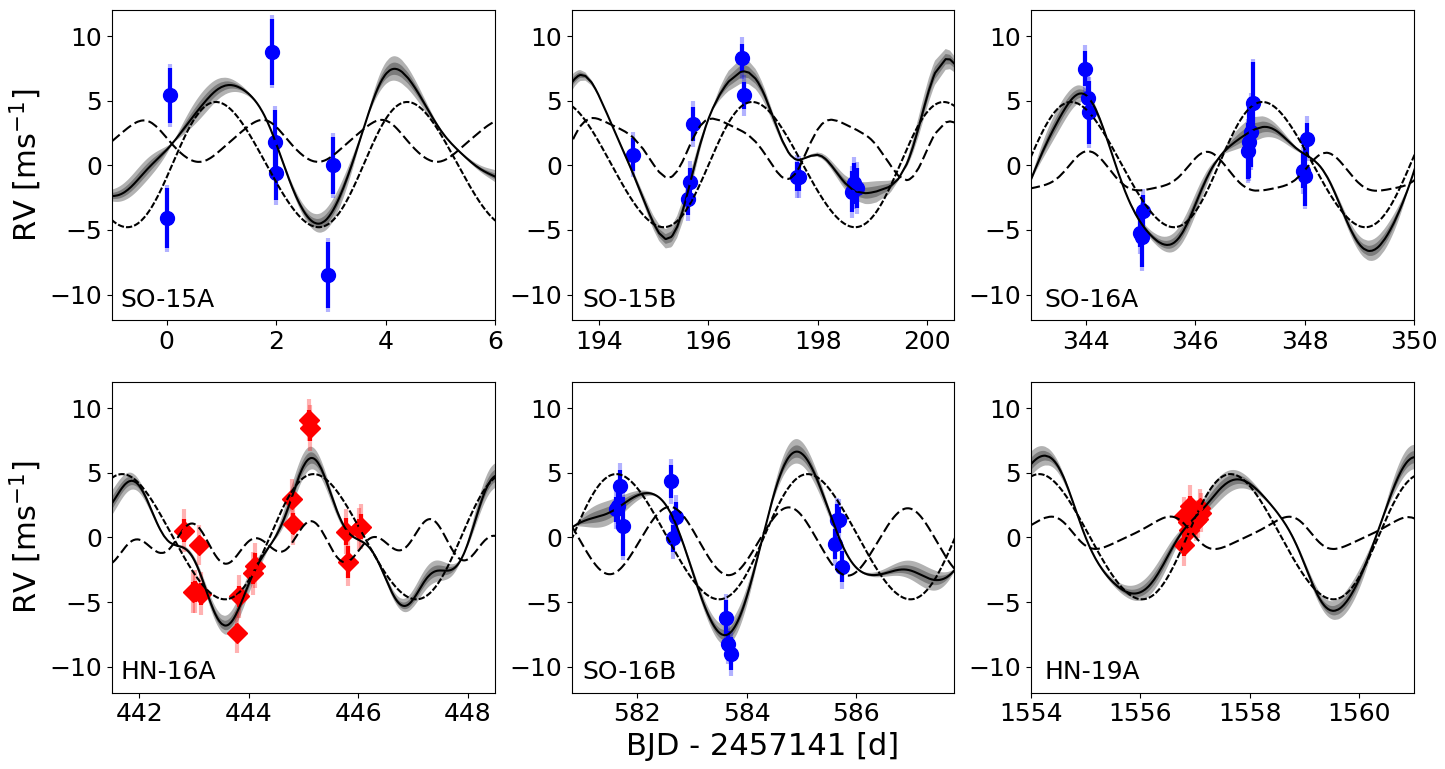}
         \label{fig:rvgp_a}
     \end{subfigure}
     \hfill
     \begin{subfigure}[b]{0.325\textwidth}
         \caption{}
         \includegraphics[width=\textwidth, trim=0mm -20mm 10mm 0mm]{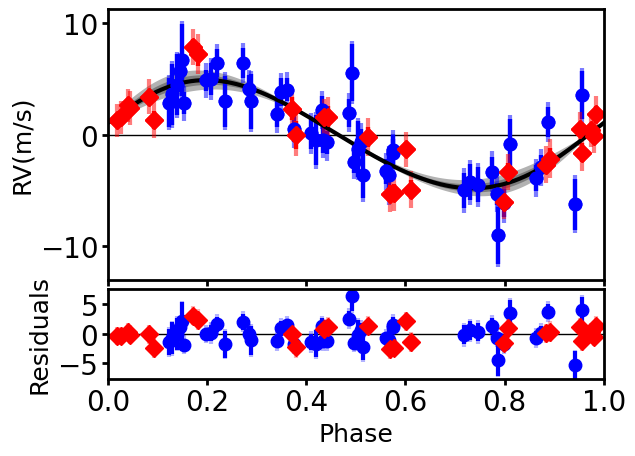}
         \label{fig:rvgp_b}
     \end{subfigure}
     \begin{subfigure}[b]{0.65\textwidth}
         \caption{}
         \includegraphics[width=\textwidth, trim=0mm 0mm 0mm 0mm]{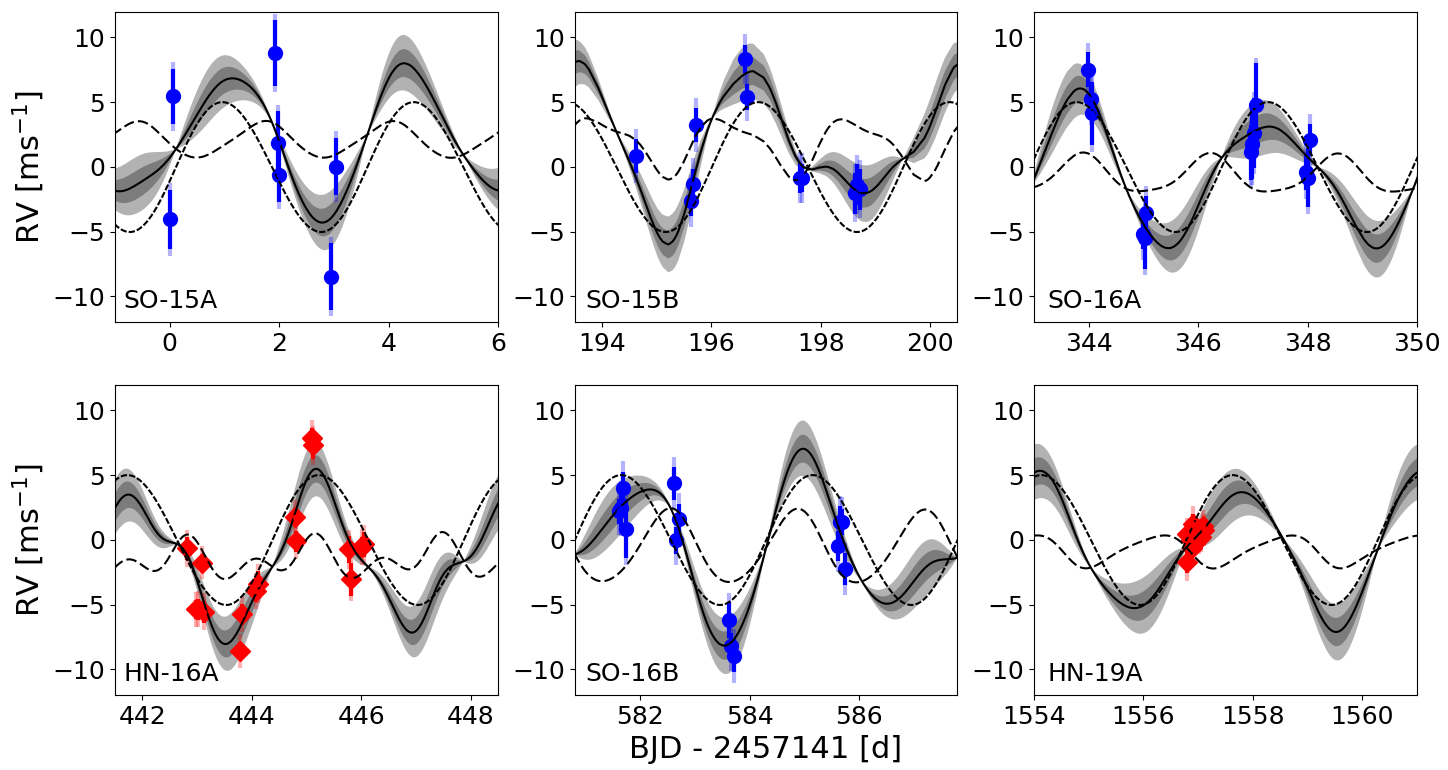}
         \label{fig:rvbisgp_a}
     \end{subfigure}
     \hfill
     \begin{subfigure}[b]{0.325\textwidth}
         \caption{}
         \includegraphics[width=\textwidth, trim=0mm -20mm 10mm 0mm]{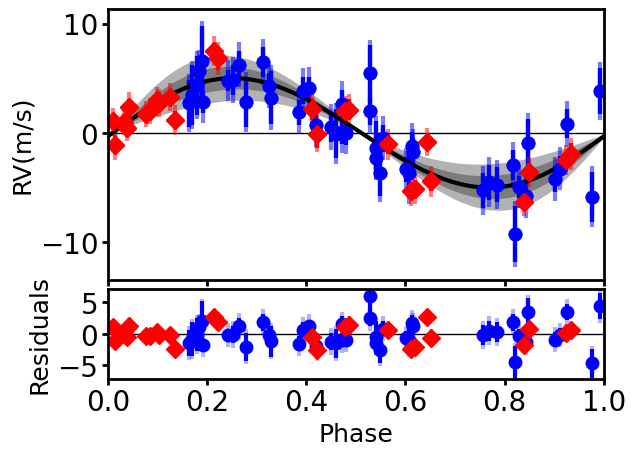}
         \label{fig:rvbisgp_b}
     \end{subfigure}
    \end{center}
     \begin{subfigure}[b]{0.65\textwidth}
         \caption{}
         \includegraphics[width=\textwidth, trim=0mm 0mm 0mm 0mm]{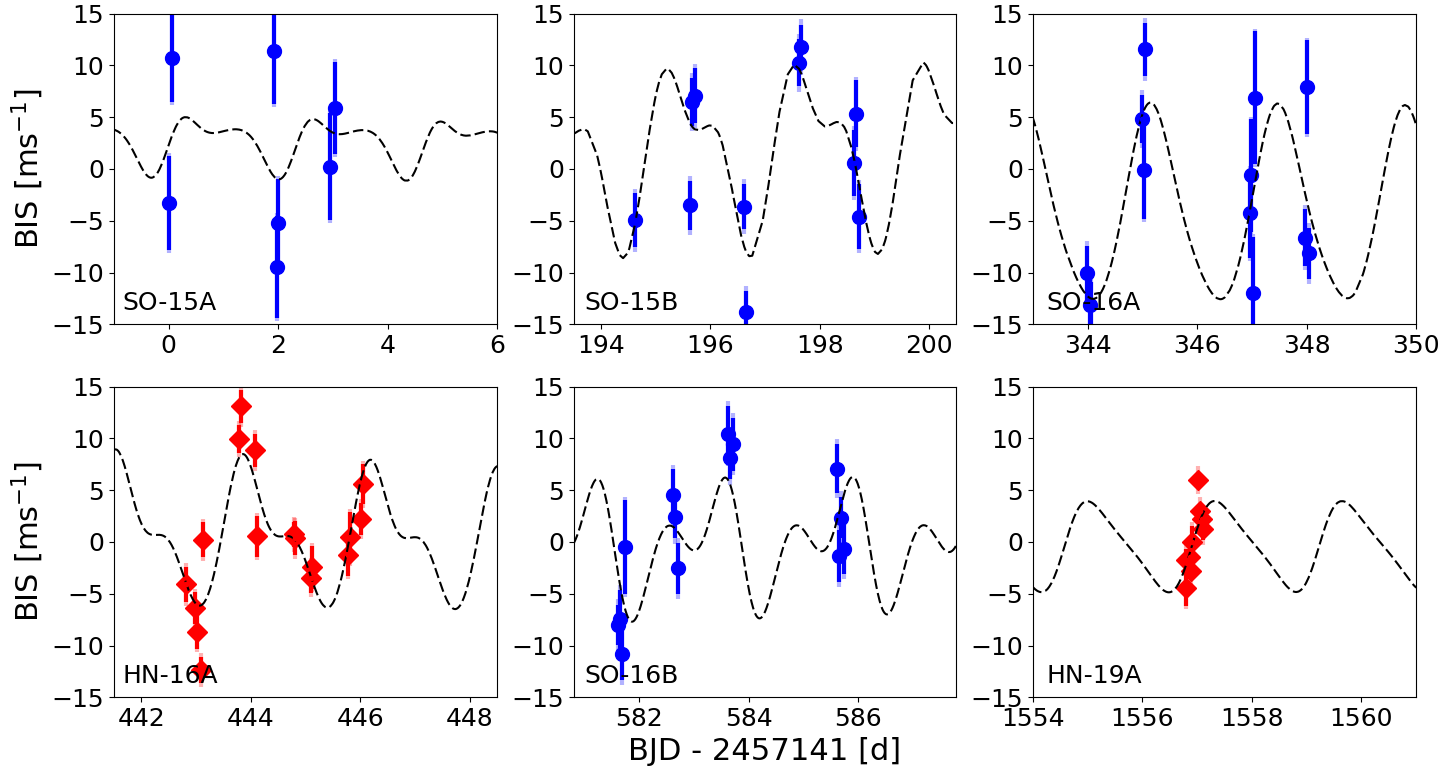}
         \label{fig:bisgp}

     \end{subfigure}
      \hfill
     \begin{subfigure}[b]{0.325\textwidth}
     \end{subfigure}

    \caption{Single Keplerian solution ($N_{\textrm{p}}=1$) with Gaussian Process (GP). (a) Solution using only the RVs with a GP. The RV component is represented by a short dashed line and the GP by a long dashed line. As in Figure 8, the RV + GP curve shaded regions show 68\% and 95\% model uncertainties. (b) The phased RVs corresponding to panel (a). (c) RVs for the RV + BIS solution with a GP and (d) corresponding phased RVs. (e) The BIS timeseries for the RV + BIS solution showing the GP as a long dashed line.}
     \label{fig:rv_1kepGP}
\end{figure*}

\section{RV analysis}
\protect\label{section:RVanalysis}

We began by assessing the coherence of periodic RV signals over time as more data points were added. Figure \ref{fig:stacked} shows the stacked Bayesian general Lomb-Scargle periodograms of the RVs and the residual RVs after accounting for the first Keplerian \citep{mortier15,mortier17}. Increasing the number of observations over the 6 observing campaigns, spanning 2015 and 2016 with the additional night in 2019 steadily increases the significance of each signal. There is some aliasing of peaks, which split into sub-peaks with addition of the final data points taken at HARPS-N in 2019. The long-term signal coherence suggests that the periodicities are Keplerian in origin and can be attributed to planets. Nevertheless, since the periodicities found in the BIS data are similar to those of the second RV signal, we investigate a number of scenarios below.

\subsection{Maximum likelihood searches}
\protect\label{section:recper_solutions}

We carried out likelihood period analyses on the combined SOPHIE and HARPS-N RVs, assuming circular Keplerian orbits with $e=0$. The Keplerian model includes offset terms for {each} instrument. In addition to a period search on the RVs only (RV model), we also performed a period search that incorporated the BIS values into the likelihood model via a linear correlation coefficient (RV+linBIScorr model). Keplerians are added recursively into the model when searching for additional signals. The likelihood model is described in detail in \cite{anglada13} and \cite{anglada16proxima} and includes white noise terms to account for noise in addition to the formal uncertainties and systematic RV offset terms for each data set.

Fig. \ref{fig:rv_periodograms} (top left) shows that periodicity at $P = 3.498$~d is found for the 1-Keplerian models ($N_\textrm{p} = 1$) for both RV and RV+linBIScorr models. Table \ref{tab:likelihood} lists the likelihood values, log$L$, and change in likelihood, $\Delta$log$L$, of recursively added periodicities. The corresponding estimates of the Bayesian Information Criteria, $\Delta\textrm{BIC} = \textrm{BIC}_{k}-\textrm{BIC}_{k-1}$ were used to estimate Bayes Factor values, $\textrm{BF} = e^{(\textrm{BIC}_{k-1}-\textrm{BIC}_{k})}/2$. The ratio of Bayesian evidence of competing models, the Bayes Factor (BF), provides a measure of support in favour of one model over another and is obtained directly from the samples obtained. We use the scale defined by \cite{trotta08bayes} and discussed in \cite{standing22bebop2}, where $3 \leq \textrm{BF} < 12$ implies weak evidence, $12 \leq \textrm{BF} < 150$ implies moderate evidence and $\textrm{BF} \ge 150$ indicates strong evidence. We note that these BF values are approximations and should not be directly compared with the values reported from posterior sampling in the following sections. For $N_\textrm{p} = 1$, $\Delta\textrm{BIC}$ for the RV+linBIScorr model is smaller than for the RV model. Nevertheless, both $\Delta\textrm{BIC}$ and BF indicate very highly significant periodicity at $P = 3.498$~d for both models.

For the 2-Keplerian solutions, with $N_\textrm{p} = 2$, significant peaks at $2.459$~d are found (Table \ref{tab:likelihood}). For the RV model, $\Delta$~Log$L = 17.3$ ($\Delta\textrm{BIC} = -21.9$). While this periodicity also contains the highest power in the RV+linBIScorr model, the significance is decreased to $\Delta$Log$L = 14.6$. The corresponding $\Delta\textrm{BIC} = -14.3$ and $\textrm{BF} = 1274$ still indicates very strong evidence for this model \citep{raftery95}, though the RV model is preferred over the RV+linBIScorr model. Although the linBIScorr correlation coefficient for HARPS-N is $-1.1\pm0.5$, the coefficient for the SOPHIE data is consistent with zero at $-0.2\pm0.7$. This suggests that the linear correlation coefficient is not adequate for modelling long baseline timeseries.

{The changing correlation between the RVs and simultaneous BIS values at different epochs, as evidenced by Pearson's $r$ statistic in Table \ref{tab:correlations}, might be indicative of an active star with continually evolving activity.
Since \dmppF~shows evidence of low level photometric modulation and evidence for BIS variability at a period that matches the signal at $\sim 2.5$~d, we investigated whether Gaussian Process (GP) modelling can be used to characterise the effects of activity on the RVs.}


\subsection{Signal recovery and model comparison using posterior sampling}
\protect\label{section:rvkep}
{We used \texttt{kima} \citep{faria18kima} to search for Keplerian signals and investigate various model scenarios. We include the maximum likelihood values in our model solutions in Table \ref{tab:bigsolution}, but note that they should not be use to distinguish models since they do not take into consideration the number of model parameters or data points. \texttt{Kima} uses Diffusive Nested Sampling \citep{brewer11diffusive} to sample from the joint posterior distribution and enables the marginal likelihood (Bayesian evidence) of a model to be directly estimated (in contrast to the BF estimates derived from the BIC values in \S \ref{section:recper_solutions}). This has the benefit of correctly enabling inter-comparison of models \citep{brewer15fast,feroz11}. With \texttt{kima}, it is possible to either fix the number of Keplerians, $N_\textrm{p}$, or include $N_\textrm{p}$ as a free parameter. The likelihood model includes the systemic velocity, $\gamma$, of a reference data set (in our case, the \hbox{HARPS-N} RV data, $\gamma_\textrm{RV,HN}$) and relative offsets for other data sets. Since our RV data are derived via matched spectrum template derived from the observations, these parameters are expected to be close to $0\,\textrm{ms}^{-1}$. As with our likelihood periodogram searches, additive white noise, $\sigma$, are included as free parameters for each data set. Further details can be found in \cite{standing22bebop2}.

\subsubsection{1-Keplerian ($N_\textrm{p}=1$) solution}\label{subsection:Np1}
\protect\label{section:rvkep1}
Using \texttt{kima} to obtain planetary parameters as described in \cite{standing22bebop2}, we find $P_\textrm{b}= 3.49733^{+0.00093}_{-0.23513}$ when we restrict the search to 1 Keplerian ($N_\textrm{p} = 1$), in agreement with the likelihood period search. The skewed uncertainties (i.e. a relatively large negative uncertainty) arise from posterior sampling of the alias peaks, which can be seen at $P=3.19$~and~$P=3.26$~d in Fig. \ref{fig:rv_periodograms} (a). The peak aliasing is a consequence of the sampling, since observations span $\leq 5$ nights at any given observing epoch with SOPHIE or HARPS-N. The $N_\textrm{p} = 1$ solution parameters are shown in Table \ref{tab:bigsolution}, column 1; the RVs and solution curve are shown in Fig. \ref{fig:rv_1kep}. We find $\textrm{BF} > 7676$ in favour of a single-planet model compared with a zero-planet model. (N.B. since no samples were obtained during the run for the $N_\textrm{p} = 0$ model, we estimate a lower limit on the BF by setting the number of samples with $N_\textrm{p} = 0$ equal to 1 \citep{standing22bebop2, triaud22bebop3, baycroft23, standing23bebop1}.

\subsubsection{2-Keplerian ($N_\textrm{p}=2$) solution}

\protect\label{section:rvkep2}
We allowed \texttt{kima} to sample posterior space with $N_\textrm{p}$ as a free parameter. In this way it is possible to recover the preferred number of Keplerians. We place a maximum $N_\textrm{p}=3$, and find the solution with $N_\textrm{p} = 2$ has the highest posterior evidence, with $\textrm{BF}=2397$, compared with the $N_\textrm{p} = 1$ model, and is shown in Fig. \ref{fig:rv_2kep}. There is not sufficient evidence for further Keplerians. However, the two preferred periods differ from the those found by the likelihood model, which adds Keplerian signals recursively. Instead of the $P_\textrm{b}=3.498$~d and $P_\textrm{c}=2.459$~d identified in \S \ref{section:recper_solutions}, the $N_\textrm{p} = 2$ model recovers $P_\textrm{b} = 2.45703^{+0.00293}_{-0.00082}$~d and $P_\textrm{c} = 5.4193^{+0.0060}_{-0.0032}$~d (Table \ref{tab:bigsolution}, column 2). With the additive white noise terms, $\sigma_\textrm{RV,SO}$ and $\sigma_\textrm{RV,HN}$, the effective uncertainties in the $N_\textrm{p}=2$ model are $1.68\,\textrm{ms}^{-1}$~and~$1.36\,\textrm{ms}^{-1}$, suggesting that the formal uncertainties, particularly for HARPS-N (see \S \ref{section:spectroscopic}), may be under-estimated.

We investigated the difference between the periods identified by the $N_\textrm{p} = 2$ recursive likelihood method in \S \ref{section:recper_solutions} and the posterior sampling method of \texttt{kima}. We performed MCMC posterior sampling on the two periods we identified in \S \ref{section:recper_solutions} and, then performed the same analysis on the periods identified by \texttt{kima}. Moderate evidence in favour of the two periodicities recovered by \texttt{kima} was found over the recursively added second periodicity, with $\Delta\textrm{BIC+}=-5.04$ and $\textrm{BF} = 12.2$. Similarly, we tried restricting the first signal to the $N_\textrm{p}=1$ period identified by \texttt{kima} and used \texttt{kima} to detect any further signals. The posterior sampling yields a second period at $2.23$~d with a moderate $\textrm{BF}=29.2$, somewhat lower than for the preferred $N_\textrm{p}=2$ solution.

The discrepancy between the periods recovered by the two approaches arises because of the data sampling, which is not optimal for recovering multiple periodicities with short observing epochs spanning only a few days. In this scenario, the multimodal posterior sampling used by \texttt{kima} may perform better than a method that adopts recursive addition of signals.

\subsection{Quasi-periodic Gaussian Process}
\protect\label{section:qpgp}

\texttt{kima} has also been adapted to enable the use of a Gaussian Process (GP). The hyperparameters of the GP are inferred together with the orbital parameters \citep{faria23kima}. The GP quasi-periodic kernel is defined in the form given by \citealt{rw06gp}(see also \citealt{haywood14}) as

\begin{equation}
\gamma_{\rm QP}(t_i,t_j) = \eta_1^2 {\rm exp} \left\{
-\frac{(t_i-t_j)^2}{2\eta_2^2}
-\frac{2}{\eta_4^2}sin^2\left[\frac{\pi(t_i-t_j)}{\eta_3}\right]
\right\}
    \protect\label{eqn:qpgp}
\end{equation}

\noindent
which describes the correlation between RVs at times $t_i$ and $t_j$. Four hyperparameters define the covariance matrix. The amplitude, $\eta_1$, describes the deviation of the GP models from the mean function. The parameter, $\eta_2$ is the long term evolution timescale, while $\eta_3$, the characteristic period of the GP, can be related directly to the stellar rotation period. As noted by \cite{rajpaul15} and \cite{barragan22pyaneti2}, it is desirable that $\eta_3 < \eta_2$, to ensure the validity of using the quasi-periodic GP with a periodic component. The factor, $\eta_4$, is proportional to the inverse of the square root of the harmonic complexity, $\Gamma$ (e.g. see \citealt{rajpaul15}, \citealt{barragan22pyaneti2} and \citealt{nicholson22gp}). From equation \ref{eqn:qpgp}, $\Gamma = 2/\eta_4^2$, and describes the roughness or complexity of variations within each characteristic period, $\eta_3$.

\subsubsection{Keplerian + quasi-periodic Gaussian Process}
\protect\label{section:rvfullgp}
Because our data comprise only a few nights per observing run, a careful choice of priors for the GP hyperparameters, is required. We tested solutions informed by the photometry and activity signatures investigated in \S \ref{section:photometry} and \S \ref{section:spectroscopic}. Following our finding of photometric periodicity (Section~\ref{section:photometry}), we first trialled a GP prior for $\eta_3$ of $\mathcal{G}[11.6\pm1.5]$~d (i.e. a Gaussian distribution).
Given the likely activity signatures at $P_{\rm rot}/2$ and $P_{\rm rot}/4$ discussed above in \S \ref{section:activityperiods}, we expect characteristic activity signatures on timescales shorter than than $P_{\rm rot}$. This suggests some degree of harmonic complexity is required to account for any structure related to the characteristic period, $\eta_3$. \cite{nicholson22gp} note, for instance, that analysis of stellar lightcurves typically yield harmonic complexity of $0.5 < \Gamma < 2$.
To investigate the posterior distribution of the harmonic complexity, we used a fairly wide prior of \hbox{0.1 < $\eta_4$ < 10}, corresponding to harmonic complexity of \hbox{$0.02 < \Gamma < 200$}.


With this configuration, we do not find evidence for any planetary signals. Although the posterior distribution returns significant periodicities at $\sim 2.5$~d, $3.5$~d~and~$5.4$~d, the posterior samples for $N_\textrm{p} = 1, 2$~and~$3$ return low respective Bayes Factors of BF = $1.2$, $1.3$ and $1.4$. We found that the prior distributions are recovered in the posteriors for $\eta_2$ and $\eta_3$. This not surprising since the few-day span of each observing epoch is much less than the observationally informed priors. The posterior distribution for $\eta_4 = 0.15^{+0.17}_{-0.04}$ corresponds to harmonic complexity of $\Gamma = 89^{+76}_{-69}$. This high harmonic complexity thus accounts for most of the RV variability. However, the sampling rate of the observations leads us to reject solutions that fit RV variability on very short timescales. There are too many fitting parameters and not enough data points to constrain a credible solution. Essentially, the model is too flexible.

\subsubsection{A simplified quasi-periodic Gaussian Process}
\protect\label{section:rvreducedgp}
The intensive sampling strategy we adopted to enable searches for shorter period planets means that several observations were made over a few hours on each night. By contrast, the significant periodicities in the data are typically a few days, with the effect that the 71 data points essentially behave as 21 high S/N data points. Since the $N_\textrm{p} = 1$~+~GP model in \S \ref {section:rvfullgp} contains 13 fitting parameters, the flexibility of the GP model and thus the degeneracy with planetary signals is high \citep{nicholson22gp}. Thus it is not surprising that the hyperparameter priors either dominate the posteriors in \S \ref {section:rvfullgp}, or lead to over-fitting when relatively uninformative and wide.

In simulations, \cite{nicholson22gp} noted that there is no single starspot parameter in their models that relates directly to $\Gamma$. They adopted initial guesses of $\Gamma = 1$ when fitting simulated lightcurves, as this value represents the transition between low and high harmonic complexity \citep{barragan22pyaneti2}. Informed by the simultaneously derived BIS periodicity of $2.54$~d, we obtained posterior samples with a prior on $\eta_3$ of $\mathcal{G}[2.5\pm0.5]$~d. In light of the posterior distribution for $\eta_4$ that we found in \S \ref{section:rvfullgp}, we followed \cite{nicholson22gp}, but {\em fixed} $\Gamma = 1$ (i.e. we fixed $\eta_4 = \sqrt{2}$). With these priors, posterior sampling leads to a preference for $N_\textrm{p} = 1$, with $P = 3.4982^{+0.0015}_{-0.0027}$~d and evidence of $\textrm{BF} = 111.4$. As in \S \ref{section:rvfullgp}, the posterior distribution is dominated by the priors for $\eta_2$, though the data appears to have had slight influence on the posterior GP periodic component $\eta_3 = 2.42^{+0.25}_{-0.24}$. The RV solution curves and folded RVS are plotted in Fig. \ref{fig:rv_1kepGP} (a\&b).
Hence, there is moderate evidence for a single Keplerian with the same period as the purely Keplerian $N_\textrm{p} = 1$ model. The posteriors show that inclusion of the GP resulted in reduced significance of the periodicities at $2.457$~d and $5.419$~d found for the purely Keplerian $N_\textrm{p} = 2$ model.

\subsection{Alternative models using activity indicators}

With the evidence for moderate activity correlations between the RV and BIS data, we used \texttt{kima} to obtain simultaneous posterior samples using both data sets and a simultaneous single GP model. Here an additional GP hyperparameter, $\eta_{1,2}$ is required to model the BIS amplitude. The number of fitting parameters for the model is 17. This is similar to the approach adopted by \cite{barragan22pyaneti2} although we do not consider the time derivative of the GP amplitude terms here.
Further, adding the BIS timeseries doubles the size of the data set to 142, or if we consider the effective data set size, to 42 as in \S \ref{section:rvreducedgp}. The flexibility of the model is thus reduced with inclusion of the activity timeseries.

\subsubsection{Model with simultaneous BIS activity indicator}

We used the same model priors as in \S \ref{section:rvfullgp}. We again found that the $\eta_2$ and $\eta_3$ posterior distributions matched the priors while a high harmonic complexity is preferred, with $\eta_4 = 0.15^{+0.05}_{-0.04}$ ($\Gamma = 89^{+76}_{-39}$).
The posterior samples demonstrate an over-density at $P \sim 3.5$~d, though for $N_\textrm{p} = 1, 2$~and~$3$, respective Bayes Factors of BF = $1.0$, $0.8$ and $0.8$ indicate very weak evidence for any Keplerian signals. Even with the inclusion of the BIS timeseries, the model is too flexible.

\subsubsection{Model with simultaneous BIS activity and fixed harmonic complexity}

Finally, with fixed $\Gamma = 1$ ($\eta_4 = \sqrt{2}$), as in \S \ref{section:rvreducedgp}, we used a prior on $\eta_3$ of $\mathcal{G}[2.5\pm0.5]$~d. The solution is shown in Table \ref{tab:bigsolution}, column 4 and in Fig. \ref{fig:rv_1kepGP} (c, d and e). A $\textrm{BF} = 1038$ indicates strong evidence in favour of a $N_\textrm{p} = 1$ solution with with $P = 3.4982$~d. The inclusion of BIS thus appears to have provided further evidence for the $2.5$~d activity periodicity, yielding a higher BF compared with the same model and priors on the RV-only timeseries in \S \ref{section:rvreducedgp}.

Fig. \ref{fig:rv_1kepGP} (e) demonstrates that there is considerable scatter in the BIS at some epochs, particularly in the poorer quality data in the SO-15A run. The white noise terms for the BIS data (Table \ref{tab:bigsolution}) reflect this. Modelling the RV and BIS data simultaneously has also likely resulted in the higher RV white noise terms compared with the  $N_\textrm{p} = 2$ solution; accounting for $\sigma_\textrm{RV,SO}$ and $\sigma_\textrm{RV,HN}$ yields respective effective uncertainties of $2.22\,\textrm{ms}^{-1}$~and~$1.48\,\textrm{ms}^{-1}$.
}

\section{Summary and Discussion}
\protect\label{section:summary}

Although there is clear evidence for a single candidate Keplerian ($N_\textrm{p}=1$) with $P_\textrm{b}~= 3.49791^{+0.00046}_{-0.00143}$~d and $m_\textrm{b}$\,sin\,$i$~= $11.6_{-1.5 }^{+1.6 }$~M$_\oplus$, the purely Keplerian model indicates strong evidence favouring a 2-Keplerian ($N_\textrm{p}=2$) solution, with a relative Bayes Factor, $\textrm{BF} = 2397$.
The preferred $N_\textrm{p}=2$ solution does not include the $N_\textrm{p}=1$ period, but rather recovers periods of $P_\textrm{b}~=~2.4570^{+0.0026}_{-0.0462}$~d and $P_\textrm{c}~=~5.4196^{+0.6766}_{-0.0030}$~d with respective derived minimum masses of $m_\textrm{b}\,\textrm{sin}\,i~= 8.0_{-1.5 }^{+1.1}$~M$_\oplus$ and $m_\textrm{c}\,\textrm{sin}\,i~= 12.2_{-1.6}^{+1.4 }$~M$_\oplus$.
The $N_\textrm{p}=2$ solution derived from likelihood periodogram searches, with $P_\textrm{b}~=~3.498$~d and $P_\textrm{c}~=~2.459$~d, suggests that there may be conditions where the common approach of recursively adding signals does not lead to the best solution; here, it is likely that the length of the candidate periods and relative length of the observing runs has resulted in ambiguity. Our finding may also in part be a limitation of the period search algorithm, which does not explore maximum likelihood space for a global maximum, but rather, iteratively adjusts the Keplerians already identified when searching for additional signals. In this respect, an algorithm capable of finding a global marginalised maximum likelihood in a potentially multimodal posterior space is to be preferred, particularly when considering multiple Keplerian signals. It is important to note that the posterior samples from \texttt{kima} also contained a significant number of samples around the $3.498$~d period (around 40\% compared with the $2.4570$~d period and 50\% compared with the $5.4196$d period).


Despite the low \hbox{log($R^{\prime}_{\rm HK}) = -5.24$} that we used to select DMPP-4 for follow-up RV observations, there is some evidence for activity-induced RV variability. This is consistent with the intrinsic stellar activity corresponding to a higher value of log($R^{\prime}_{\rm HK}$) than observed. The hypothesis underlying DMPP is that circumstellar absorption depresses the log($R^{\prime}_{\rm HK}$), so this is as expected. Because the $N_\textrm{p}=2$ solution recovers periods that are close to one half and one quarter integer factors of the inferred rotation period ($P_\textrm{rot}/2$ and $P_\textrm{rot}/4$), as discussed in detail in \S \ref{section:activityperiods}, we were prompted to investigate RV models that contained an activity component.

The GP models we investigated for \dmppF~were necessarily restrictive because of the nature of the data sampling, the relatively short observing runs and the effective size of the data sets. When a quasi-periodic GP is included in the model, we find moderate evidence ($\textrm{BF} = 111.4$) for a single planet in the RVs with $P_\textrm{b}~= 3.4982^{+0.0015}_{-0.0027}$~d and $m_\textrm{b}\,\textrm{sin}\,i~= 12.6^{+1.6}_{-1.8} \textrm{M}_\oplus$. Including the simultaneous BIS measurements augments the Bayesian evidence, yielding $P_\textrm{b}~= 3.4982^{+0.0022}_{-0.2327}$~d and $m_\textrm{b}\,\textrm{sin}\,i~= 12.2^{+1.8}_{-1.9} \textrm{M}_\oplus$ with a $\textrm{BF} = 1038$.

Thus, although the purely Keplerian model prefers $N_\textrm{p}=2$, the model restricted to $N_\textrm{p}=1$ and the models with a GP and no $N_\textrm{p}$ restriction all provide moderate to strong evidence for a single planet. The periods and minimum masses are consistent within the uncertainties for these models. For the Keplerian models with a GP, with and without simultaneous BIS data, either greater model flexibility, the adopted tighter priors, or both, result in lower overall Bayesian evidence. Without further data it is thus difficult to distinguish between the preferred $N_\textrm{p}=2$ model and models that return single Keplerians in the presence of variable activity. Nevertheless, there is (i) evidence for activity periodicities that coincide with the identified RV periodicities in the purely Keplerian $N_\textrm{p}=2$ model. Further, there is (ii) a not-insignificant appearance of the the $3.498$~d periodicity in the model posterior samples for the $N_\textrm{p}=2$ model. And finally, (iii) a preferred periodicity at $3.4982$~d is found in the Keplerian + GP models. These three observations might be taken as reasonable evidence to prefer the solutions with a single $3.498$~d Keplerian.

We based our choices of GP hyperparameter priors on starspot distribution hypotheses and photometrically inferred periodicities. Recently however, \cite{nicholson22gp} have suggested that harmonic complexity priors derived from photometry are not necessarily appropriate. Simplifying the model by removing the harmonic complexity, $\Gamma$, and relying only on simultaneously derived and modelled BIS periodicities makes intuitive sense, especially in light of the relatively few available effective data points. \cite{nicholson22gp} also investigated the effect of reducing the number of data points and hence increasing the flexibility of the model. The resulting increased degeneracy highlights the need for sufficient data points.

The following sections further investigate the nature of the planet candidates. Despite a preference for the models with a single planet, we also considered the the $N_\textrm{p}=2$ solution.

\begin{figure}
	\begin{center}
      \includegraphics[trim=0mm 5mm 0mm 0mm, width=1\columnwidth]{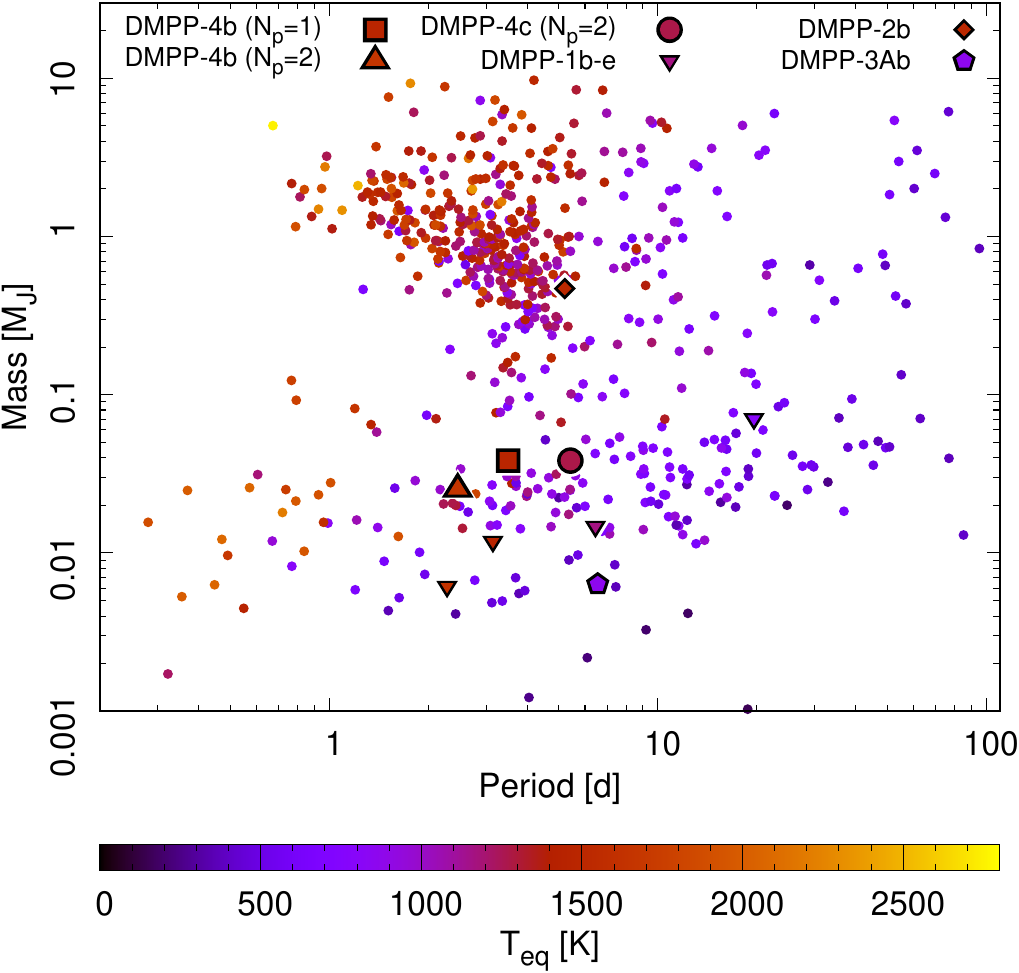} \\
	\end{center}
    \caption{The Planet mass vs orbital period diagram for planets with known mass. Each planet is colour coded according to its estimated equilibrium temperature, $T_{\rm eq}$ (assuming a Bond Albedo of $A_{\rm B} = 0.36$). {The preferred $N_\textrm{p}=1$ solution in column 4 of Table \ref{tab:bigsolution} for \dmppF~b is shown by the large square symbol. The $N_\textrm{p}=2$ candidate planets, \dmppF~b and~c, are shown by the large upright triangle and circle}. DMPP-1~b,\,c,\,d\,\&\,e, DMPP-2~b and DMPP-3A~b are also highlighted.}
    \label{fig:mass_vs_porb}
\end{figure}

\subsection{Orbital stability of planets in the \texorpdfstring{$N_{\textrm{p}}=2$}{Lg}~model}

\protect\label{section:properties_stability}

We checked the stability of the orbital solution for the {$N_\textrm{p} = 2$ model} using the IAS15 integrator within the N-body orbital integrator, {\sc rebound} \citep{rein12rebound,rein2015ias15}. {Simulations were started with $e=0$ and the semi-major axes listed in Table \ref{tab:bigsolution} for the $N_\textrm{p} = 2$ solution. For both planet candidates, only very small fluctuations in $a$ and $e$ are seen. For \dmppF~b, we find semi-major axis r.m.s. variability over 25,000 years of $5.5\times10^{-7}$~AU (i.e. 0.0014\% of the semi-major axis). A very low degree of eccentricity of $e_{\rm b}  = 0.00020 \pm 0.00009$ is established. For the putative \dmppF~c, we find corresponding r.m.s. semi-major axis variability of $4.7\times10^{-7}$~AU (i.e. 0.0007\% of the semi-major axis) and eccentricity $e_{\rm c}  = 0.00007 \pm 0.00003$. For planets that formed in circular orbits, or evolved into non-eccentric orbits via steady tidal circularisation, we expect a stable configuration. Guided by the upper limit to the eccentricities we found in Table \ref{tab:bigsolution}, we conducted simulations that started with $e=0.065$. The respective semi-major axis r.m.s. variabilities for \dmppF~b and \dmppF~c are 0.0023\% and 0.0019\% and the corresponding eccentricities were $e_\textrm{b} = 0.071 \pm 0.030$ with a full range of $0.017 < e_\textrm{b} < 0.107$ and $e_\textrm{c} = 0.055 \pm 0.019)$ with a range of $0.024 < e_\textrm{c} < 0.079$. Apart from the $680$~year periodic eccentricity variability, the orbits did not show any further evidence of evolution on the 25,000 year timescale of the simulations. The stability of the orbits is not surprising given that the closest approach of the planets is $0.021$~AU, while the respective Hill radii of \dmppF~b and~c are $r_H = 0.0007$~AU and $0.0013$~AU.


}

\subsection{Fundamental properties of the exoplanet candidates}
\protect\label{section:properties_demographics}

Fig. \ref{fig:mass_vs_porb} shows the period-mass diagram for planets with known mass {determined with $\leq$\,$20\%$ uncertainty. Only archival planets\footnote{https://exoplanetarchive.ipac.caltech.edu} with true dynamical mass estimates and tabulated $T_\textrm{eff}$ and $R_*$ are plotted (e.g. planets with lower mass estimates, $m_\textrm{p}\,\textrm{sin}\,i$, are excluded). The $N_\textrm{p}=1$, \dmppF~b, solution using RV and BIS data (Table \ref{tab:bigsolution}, column 4) and the $N_\textrm{p}=2$ solution (Table \ref{tab:bigsolution}, column 2) are shown.} The planet candidates are closer to the Neptune desert than the planets orbiting DMPP-1, -2 and -3 \citep{staab20dmpp1,haswell20dmpp,barnes20dmpp3}. {All DMPP planets possess lower mass estimates to better than 20\% with the exception of DMPP-1 e, which has an uncertainty of 22\%. The points in Fig. \ref{fig:mass_vs_porb} are} colour coded according to the planet equilibrium temperatures, $T_{\rm eq}$. The value obtained depends on the assumed albedo. In the Solar System Bond albedos, $A_{\rm B}$ vary between $\sim 0.088$ for Mercury \citep{mallama17mercurybondalbedo}, through a mean value of $0.36$ for gas giant planets (\citealt{li18jupiterbondalbedo,hanel83saturnbondalbedo,pearl90uranusbondalbedo,pearl91neptunebondalbedo}) up to $0.76$ for Venus \citep{haus16venusbondalbedo}. Using these values in the equation
$T_{\rm eq} = T_{\rm eff}\sqrt{R_*/2a}(1-A_{\rm B})^{0.25}$ therefore gives a range of plausible equilibrium temperatures.  {For the $N_\textrm{p}=1$, respective $T_{\rm eq} \sim 1608$, $1472$ and $1152$~K are implied for \dmppF~b. For $N_\textrm{p}=2$, the corresponding temperatures for \dmppF~b are $T_{\rm eq} \sim 1809$, $1655$ and $1295$~K and for  \dmppF~c are $T_{\rm eq} \sim 1388$, $1270$ and $994$~K. The equilibrium temperatures for the Mercury-like, low albedo cases are thus around $400-500$~K higher.

Using the probabilistic {\sc Forecaster} \citep{chen17forecaster} and the minimum mass for our preferred $N_\textrm{p}=1$ solution (Table \ref{tab:bigsolution}), we estimate the minimum radius of \dmppF~b as $R_{\rm b} = 3.49^{+1.47}_{-1.04}\, \rm{R}_{\oplus}$. For the $N_\textrm{p}=1$ solution, minimum radii of $R_{\rm b} = 2.72^{+1.17}_{-0.81}\, \rm{R}_{\oplus}$ and $R_{\rm c} = 3.49^{+1.46}_{-1.02}\, \rm{R}_\oplus$ are predicted.} \dmppF~b and~c {would thus be} likely to possess few per cent $H_2$ dominated atmospheres \citep{zeng19}. The planets may have lost much of their atmospheres, through radiation-driven mass loss, evolving downwards through the Neptune Desert (Fig.~\ref{fig:mass_vs_porb}). Given this probable evolutionary history, and the ongoing high irradiation from the F7V host star, \dmppF~b and~c may deviate from expectations based on the general planet population. If \dmppF~b and \dmppF~c are rocky `Chthonian' planets stripped of their atmospheres, they are likely hot enough that their surfaces comprise liquid magma oceans. Temperatures of 1100 – 1500 K are typical for molten magma on the Earth \citep{Sigurdsson15volcanoes}, as noted by \cite{staab20dmpp1}. This more extreme scenario could mean that \dmppF~b and~c lie below the exoplanet radius valley, with radii of less than $1.7\,-\,2$~R$_\oplus$ \citep{fulton17gap,zeng21gap}{, in turn implying densities of between $1.6 \times \rho_\textrm{Earth}$ and $2.5 \times \rho_\textrm{Earth}$ if $R \leq 1.7$~R$_\oplus$. This could potentially affect mass loss since the surface gravities would be $2.8-4.2$ greater, but needs detailed modelling to quantify. If substantial mass-loss does not occur, and the planets are indeed atmosphereless, to satisfy the DMPP hypothesis, we might expect that there are additional interior and as yet undetected mass-losing planets in the DMPP-4 system.}

Direct size determinations from transits would reveal bulk properties and potential atmospheric compositions. If the rotation period of \dmppF~is 11.6~d and the orbital angular momentum of the planets is aligned with the stellar rotational angular momentum,
there is a high transit probability \citep{haswell20dmpp}.  We used the Lightcurve Analysis Tool for Transiting Exoplanets ({\sc latte}) to search for transits in the 2 minute cadence {\sc tess} photometry \citep{eisner20latte}. {\sc latte} corrects the data for residual systematics using an iterative non-linear filter \citep{aigrain04planetprospecting}. A Box-Least-Squares period search is performed on the flattened lightcurves. We do not find evidence for transits in any of the {\sc tess} sectors (see \S \ref{section:photometry}) observed to date. Of course these planets may not be aligned exactly edge-on, so may not transit. In this case information from phase curves, e.g. using JWST could be revealing.


\section{Conclusion}
\protect\label{section:conclusion}

Because it is so bright, \dmppF~is an important target for follow-up characterisation. Although the current data do not enable us to unambiguously distinguish between a system with one or two planets, there is significant evidence to favour models with a single planet. Our currently preferred solution indicates a $12.2~\textrm{M}_\oplus$ planet in a $3.498$~d orbit. The potential activity contributions, amplitude modulations and periods of the \dmppF~RV signals we observe indicate that an intensive RV monitoring campaign is needed. This would better enable us to distinguish between multi-planet solutions and activity signals on the timescales of the stellar and planet candidate periods. It may also reveal additional planets below our current mass detection threshold.

\section*{Acknowledgements}

J.R. Barnes, M.R. Standing and C.A. Haswell were funded by STFC under consolidated grant ST/T000295/1 and ST/X001164/1. The radial velocity data were obtained at through The OPTICON Trans-National Access Programme (FP7II, 2013-2016). This work also used observations from the Las Cumbres Observatory global telescope network. This paper includes data collected by the TESS mission. Funding for the TESS mission is provided by the NASA's Science Mission Directorate. Data from the European Space Agency (ESA) mission {\it Gaia} (\url{https://www.cosmos.esa.int/gaia}) have been used. The {\it Gaia} data were processed by the {\it Gaia} Data Processing and Analysis Consortium (DPAC, \url{https://www.cosmos.esa.int/web/gaia/dpac/consortium}). Funding for the DPAC has been provided by national institutions, in particular the institutions participating in the {\it Gaia} Multilateral Agreement. This research made use of the NASA Exoplanet Archive, which is operated by the California Institute of Technology, under contract with the National Aeronautics and Space Administration under the Exoplanet Exploration Program (see \url{https://exoplanetarchive.ipac.caltech.edu} for further details of data sources).

\section*{Data Availability}
The data underlying this article will be shared on reasonable request to the corresponding author. The reduced radial velocities and activity indices are tabulated in Appendix A.



\bibliographystyle{mnras}
\bibliography{master,ownrefs}




\appendix




\section{Data}

\begin{table*}

\caption{SOPHIE RVs and activity indices {with their respective uncertainties}.}
    \begin{tabular}{ccccccccccc}
    \hline
    BJD        &       RV      &    $\sigma_{\textrm{RV}}$         &      $\kappa$ &  $\sigma_\kappa$  &     FWHM      &      $\sigma_{\textrm{FWHM}}$  &      BIS      &       $\sigma_{\textrm{BIS}}$  &     Sindex    &      $\sigma_{\textrm{Sindex}}$ \\

    -2450000 &   [ms$^{-1}$]   &   [ms$^{-1}$]   &               &               &     [ms$^{-1}$] &     [ms$^{-1}$] &   [ms$^{-1}$]   &  [ms$^{-1}$]    &               &                \\
    \hline

 7141.601584 &      -4.39524 &       2.28894 &      -0.17840 &       0.04610 &      -4.72111 &       2.25224 &      -3.28979 &       2.25224 &       0.00344 &       0.00057  \\
 7141.653250 &       5.10336 &       2.13098 &      -0.22447 &       0.04449 &      10.93894 &       1.91173 &      10.71020 &       1.91173 &       0.00123 &       0.00047  \\
 7143.531089 &       8.48712 &       2.55042 &       1.01775 &       0.07311 &      92.78629 &       2.45182 &      11.35026 &       2.45182 &       0.01114 &       0.00081  \\
 7143.573198 &       1.51026 &       2.47348 &      -0.01728 &       0.03295 &      16.32131 &       2.50031 &      -9.45646 &       2.50031 &       0.00367 &       0.00049  \\
 7143.602950 &      -0.92205 &       2.11206 &      -0.27088 &       0.02228 &      18.29004 &       1.91026 &      -5.22389 &       1.91026 &       0.00225 &       0.00069  \\
 7144.557581 &      -8.80357 &       2.58044 &       0.59530 &       0.08676 &      53.22427 &       3.01222 &       0.21021 &       3.01222 &       0.01168 &       0.00094  \\
 7144.646502 &      -0.32047 &       2.22101 &      -0.58017 &       0.04569 &     -21.83060 &       2.29269 &       5.87688 &       2.29269 &       0.00260 &       0.00056  \\
 7336.228379 &       0.51648 &       1.29009 &       0.24119 &       0.02395 &     -11.59571 &       1.61949 &      -4.94574 &       1.61949 &      -0.00402 &       0.00040  \\
 7337.233243 &      -2.94198 &       1.18561 &      -0.12274 &       0.02887 &     -42.31921 &       1.35967 &      -3.51409 &       1.35967 &      -0.00274 &       0.00033  \\
 7337.268115 &      -1.63685 &       1.15477 &      -0.07434 &       0.02799 &     -25.78468 &       1.35227 &       6.47557 &       1.35227 &      -0.00374 &       0.00034  \\
 7337.319238 &       2.88576 &       1.33344 &       0.93347 &       0.07464 &      48.30336 &       1.33257 &       7.05742 &       1.33257 &       0.00457 &       0.00049  \\
 7338.222375 &       8.00376 &       1.07203 &       0.31575 &       0.02922 &     -12.68507 &       1.35549 &      -3.62720 &       1.35549 &      -0.00288 &       0.00034  \\
 7338.250928 &       5.09088 &       1.00829 &       0.56340 &       0.03536 &       2.26548 &       1.35064 &     -13.84313 &       1.35064 &      -0.00259 &       0.00036  \\
 7339.228004 &      -1.18240 &       1.13805 &      -0.04812 &       0.02567 &     -11.14233 &       1.23464 &      10.23551 &       1.23464 &      -0.00143 &       0.00031  \\
 7339.257860 &      -1.20310 &       1.07228 &      -0.11327 &       0.02462 &     -15.81676 &       1.24214 &      11.78683 &       1.24214 &      -0.00263 &       0.00030  \\
 7340.227075 &      -2.35796 &       1.58780 &       0.46295 &       0.03608 &     -22.86128 &       1.65094 &       0.54354 &       1.65094 &      -0.00113 &       0.00042  \\
 7340.263479 &      -1.70372 &       1.59813 &       0.48781 &       0.03798 &     -17.09651 &       1.76456 &       5.34441 &       1.76456 &      -0.00339 &       0.00046  \\
 7340.319923 &      -2.06757 &       1.57236 &       2.25613 &       0.10606 &      45.08994 &       1.87082 &      -4.61932 &       1.87082 &       0.00246 &       0.00064  \\
 7485.596614 &       7.17563 &       1.33677 &       1.39810 &       0.07858 &      53.23429 &       1.31924 &     -10.08458 &       1.31924 &       0.00737 &       0.00044  \\
 7485.637771 &       4.89580 &       1.21995 &       0.45280 &       0.03302 &      31.25865 &       1.15977 &     -13.16582 &       1.15977 &       0.00284 &       0.00033  \\
 7485.653947 &       3.79001 &       2.45915 &      -0.28637 &       0.04497 &     -16.98751 &       2.33639 &     -15.78980 &       2.33639 &      -0.00280 &       0.00057  \\
 7486.602113 &      -5.52468 &       1.14257 &       1.00287 &       0.05604 &      55.68851 &       1.25160 &       4.81432 &       1.25160 &       0.00206 &       0.00036  \\
 7486.620842 &      -5.85710 &       2.34468 &       0.89867 &       0.09027 &      40.23651 &       2.11967 &      -0.12312 &       2.11967 &       0.00115 &       0.00060  \\
 7486.651285 &      -3.88557 &       1.26869 &       0.29901 &       0.03039 &      15.83827 &       1.23164 &      11.53410 &       1.23164 &      -0.00099 &       0.00031  \\
 7488.568901 &       0.77768 &       2.15324 &       1.65057 &       0.17660 &      59.88006 &       2.28861 &      -4.28979 &       2.28861 &       0.00672 &       0.00077  \\
 7488.589914 &       1.46278 &       2.71892 &       0.69793 &       0.10059 &      34.68188 &       2.33840 &      -0.62312 &       2.33840 &      -0.00030 &       0.00066  \\
 7488.628896 &       2.29283 &       2.69899 &       0.92556 &       0.10769 &      54.12248 &       3.48791 &     -11.95650 &       3.48791 &      -0.00659 &       0.00099  \\
 7488.660901 &       4.47383 &       3.22298 &       0.06100 &       0.05604 &      -2.11315 &       3.23877 &       6.87688 &       3.23877 &      -0.00663 &       0.00081  \\
 7489.572033 &      -0.72295 &       1.36631 &       2.11115 &       0.08152 &      73.07962 &       1.29323 &      -6.62897 &       1.29323 &       0.01013 &       0.00046  \\
 7489.605550 &      -1.16558 &       2.26300 &       1.52788 &       0.13017 &      49.23823 &       2.24038 &       7.87688 &       2.24038 &       0.00647 &       0.00072  \\
 7489.647550 &       1.75240 &       1.23957 &       0.16331 &       0.03525 &      -3.23986 &       1.55964 &      -8.13880 &       1.55964 &      -0.00186 &       0.00040  \\
 7723.218982 &       1.87889 &       0.96311 &      -0.41169 &       0.02352 &     -42.80423 &       1.04460 &      -8.06201 &       1.04460 &      -0.00316 &       0.00024  \\
 7723.253581 &       2.09855 &       1.40480 &      -0.16232 &       0.03324 &     -29.67417 &       1.34340 &      -7.40832 &       1.34340 &      -0.00076 &       0.00033  \\
 7723.295186 &       3.68135 &       1.25362 &       0.06011 &       0.03636 &     -24.18203 &       1.32260 &     -10.86468 &       1.32260 &      -0.00045 &       0.00034  \\
 7723.350564 &       0.53657 &       2.27489 &       0.03555 &       0.04458 &     -12.12334 &       2.25218 &      -0.45646 &       2.25218 &       0.00014 &       0.00058  \\
 7724.224800 &       4.01791 &       1.25577 &      -0.49720 &       0.03038 &     -34.87339 &       1.34514 &       4.48724 &       1.34514 &      -0.00296 &       0.00032  \\
 7724.261675 &      -0.38338 &       1.04110 &      -0.02925 &       0.02706 &     -20.21665 &       1.03263 &       2.43856 &       1.03263 &      -0.00084 &       0.00026  \\
 7724.313087 &       1.24111 &       1.22726 &      -0.06321 &       0.02821 &     -28.89020 &       1.31397 &      -2.56293 &       1.31397 &      -0.00118 &       0.00033  \\
 7725.233619 &      -6.52162 &       1.35877 &      -0.20534 &       0.02654 &     -16.29436 &       1.33851 &      10.43826 &       1.33851 &      -0.00185 &       0.00033  \\
 7725.270120 &      -8.54515 &       0.99990 &      -0.18854 &       0.02127 &     -15.14537 &       1.14938 &       8.07956 &       1.14938 &      -0.00152 &       0.00029  \\
 7725.323129 &      -9.30088 &       1.27309 &       0.22961 &       0.02543 &       2.19168 &       1.29251 &       9.42790 &       1.29251 &       0.00006 &       0.00034  \\
 7727.215701 &      -0.80584 &       1.17479 &      -0.25188 &       0.02333 &     -32.46199 &       1.15532 &       7.07822 &       1.15532 &      -0.00319 &       0.00028  \\
 7727.247700 &       1.01686 &       1.25185 &      -0.07080 &       0.02174 &     -16.33009 &       1.14465 &      -1.37031 &       1.14465 &      -0.00205 &       0.00029  \\
 7727.291796 &       1.05599 &       1.00530 &       0.03403 &       0.02116 &     -24.66017 &       1.30235 &       2.35188 &       1.30235 &      -0.00232 &       0.00033  \\
 7727.339162 &      -2.57386 &       1.21831 &       0.78096 &       0.04290 &      11.39416 &       1.26968 &      -0.66842 &       1.26968 &       0.00350 &       0.00038  \\
    \end{tabular}
\label{tab:sophie_rvs_and_activity}
\end{table*}

\begin{table*}
\caption{HARPS-N RVs and activity indices {with their respective uncertainties}.}
    \begin{tabular}{ccccccccccc}
    \hline
BJD        &       RV      &    $\sigma_{\textrm{RV}}$         &      $\kappa$ &  $\sigma_\kappa$  &     FWHM      &      $\sigma_{\textrm{FWHM}}$  &      BIS      &       $\sigma_{\textrm{BIS}}$  &     Sindex    &      $\sigma_{\textrm{Sindex}}$ \\

    -2450000 &   [ms$^{-1}$]   &   [ms$^{-1}$]   &               &               &     [ms$^{-1}$] &     [ms$^{-1}$] &   [ms$^{-1}$]   &  [ms$^{-1}$]    &               &                \\
    \hline


 7584.425790 &      -0.08931 &       0.86354 &      -0.67475 &       0.04260 &      -9.42621 &       0.85020 &      -4.10532 &       0.85020 &      -0.00065 &       0.00025  \\
 7584.582598 &      -4.81271 &       0.75718 &      -0.01508 &       0.01218 &       0.21320 &       0.74623 &      -6.39534 &       0.74623 &      -0.00036 &       0.00021  \\
 7584.613142 &      -4.79456 &       0.80999 &      -0.75987 &       0.02489 &      -0.89548 &       0.81160 &      -8.69211 &       0.81160 &      -0.00054 &       0.00023  \\
 7584.700103 &      -1.18472 &       0.61352 &      -2.71648 &       0.06898 &       1.59290 &       0.62110 &     -12.34777 &       0.62110 &      -0.00156 &       0.00018  \\
 7584.729063 &      -4.98745 &       0.85418 &      -1.62125 &       0.01980 &       5.90301 &       0.69821 &       0.21189 &       0.69821 &      -0.00121 &       0.00021  \\
 7585.385055 &      -7.99774 &       0.66480 &       2.79469 &       0.02110 &      14.57599 &       0.65326 &       9.93206 &       0.65326 &       0.00010 &       0.00019  \\
 7585.415764 &      -5.17417 &       0.80530 &       2.55094 &       0.02791 &       7.35933 &       0.76919 &      13.07475 &       0.76919 &      -0.00008 &       0.00021  \\
 7585.680900 &      -3.39515 &       0.80822 &       7.07734 &       0.19464 &      -5.43576 &       0.85222 &       8.84914 &       0.85222 &       0.00258 &       0.00024  \\
 7585.710733 &      -2.81989 &       0.98120 &       6.11499 &       0.12538 &      -1.54140 &       0.88054 &       0.54579 &       0.88054 &       0.00115 &       0.00025  \\
 7586.385670 &       2.35233 &       0.61816 &       2.99824 &       0.03486 &      -6.74626 &       0.63072 &       0.73490 &       0.63072 &       0.00082 &       0.00018  \\
 7586.416934 &       0.45226 &       0.81986 &       2.90381 &       0.07114 &      -7.84784 &       0.86834 &       0.34820 &       0.86834 &       0.00077 &       0.00024  \\
 7586.695960 &       8.46464 &       0.74669 &       6.98437 &       0.20156 &      -5.80447 &       0.72816 &      -3.43951 &       0.72816 &       0.00193 &       0.00020  \\
 7586.725758 &       7.86117 &       1.02063 &       7.11873 &       0.27284 &       7.95727 &       0.97756 &      -2.46851 &       0.97756 &       0.00112 &       0.00028  \\
 7587.387682 &      -0.17318 &       1.02249 &       3.76661 &       0.10800 &      18.76964 &       0.95000 &      -1.25455 &       0.95000 &       0.00108 &       0.00028  \\
 7587.421027 &      -2.49737 &       1.24193 &       4.96372 &       0.17343 &      14.83074 &       1.11321 &       0.43501 &       1.11321 &       0.00080 &       0.00032  \\
 7587.617955 &       0.02368 &       0.78475 &       4.22303 &       0.10651 &      13.25287 &       0.73595 &       2.22708 &       0.73595 &       0.00007 &       0.00020  \\
 7587.647257 &       0.21431 &       0.96391 &       4.62988 &       0.12810 &      12.59398 &       0.86395 &       5.54800 &       0.86395 &       0.00096 &       0.00023  \\
 8698.388858 &       1.01422 &       0.51697 &       2.76669 &       0.04796 &      20.75482 &       0.45840 &      -1.71191 &       0.45840 &       0.00022 &       0.00013  \\
 8698.404002 &      -1.17661 &       0.84289 &       0.85213 &       0.10445 &      22.52849 &       0.74091 &      -4.47112 &       0.74091 &      -0.00048 &       0.00020  \\
 8698.464088 &       0.57934 &       0.38104 &       1.41506 &       0.05223 &     -15.96659 &       0.44190 &      -1.44664 &       0.44190 &      -0.00049 &       0.00011  \\
 8698.486873 &      -0.06746 &       0.44804 &       2.37282 &       0.04728 &     -11.83323 &       0.52289 &      -2.76537 &       0.52289 &      -0.00043 &       0.00014  \\
 8698.500097 &       1.79272 &       0.77203 &      -1.51143 &       0.22837 &     -30.51439 &       1.31841 &       0.01580 &       1.31841 &      -0.00162 &       0.00032  \\
 8698.623381 &       0.61839 &       0.35713 &      -1.20359 &       0.07589 &     -18.28953 &       0.45398 &       5.98544 &       0.45398 &      -0.00115 &       0.00011  \\
 8698.646135 &       0.78231 &       0.35929 &      -0.98089 &       0.07068 &     -13.75577 &       0.42110 &       2.96804 &       0.42110 &      -0.00084 &       0.00010  \\
 8698.697318 &       1.64556 &       0.38615 &       2.33327 &       0.03592 &       7.44778 &       0.39674 &       2.23658 &       0.39674 &      -0.00008 &       0.00011  \\
 8698.714388 &       1.30359 &       0.52988 &       3.74413 &       0.05086 &      15.54402 &       0.51448 &       1.23481 &       0.51448 &       0.00046 &       0.00015  \\
    \end{tabular}
\label{tab:harps_rvs_and_activity}
\end{table*}


\bsp	
\label{lastpage}
\end{document}